\documentclass[a4paper,11pt]{article}
\pdfoutput=1 

\usepackage{jinstpub} 
\usepackage{ifpdf}
\usepackage{hyperref}
\usepackage{lineno}
\usepackage{subcaption}
\usepackage{amsmath}
\usepackage{xcolor}
\usepackage{siunitx}
\usepackage{graphicx}
\usepackage{textcomp}

\usepackage{float}
\usepackage{array, booktabs, adjustbox, makecell}
\usepackage{array}
\usepackage [english]{babel}
\usepackage [autostyle, english = american]{csquotes}

\MakeOuterQuote{"}

\title{\boldmath Recovery of coincident frequency domain multiplexed detector pulses using sequential deconvolution}

\author[a,1]{M. Mishra,\note{Corresponding author.}}
\author[a]{J. Mattingly}

\affiliation[a]{Department of Nuclear Engineering,\\ North Carolina State University, \\Raleigh, NC 27695}

\emailAdd{mmishra4@ncsu.edu}

\abstract{Multiplexing of radiation detector signals into a single channel significantly reduces the need for a large number of digitizer channels, which reduces the cost and the power consumption of a data acquisition system. We previously demonstrated frequency domain multiplexing by convolution using a prototype system that multiplexed two EJ-309 organic scintillators signals into a single channel. Each detector pulse was converted to a damped sinusoid which was then combined into a single channel. The combined signal was digitized and the original detector signal was recovered from the damped sinusoid by deconvolution. 
In this paper, we demonstrate the recovery of multiple detector signals that arrive during the same digitized record via a new sequential deconvolution method. When two detectors produce signals in the same digitized record and their pulses do not overlap in time, we found that the charge, arrival time, and particle type can be estimated fairly precisely for the first pulse, but the second pulse exhibits substantial degradation in the precision of the estimated charge and arrival time. When the pulses overlap in time, we demonstrate both theoretically and experimentally that the part of the first pulse that does not overlap with the second can be recovered accurately, so the arrival time and amplitude of the first pulse can be estimated fairly precisely, but not the charge or particle type. None of these quantities can be estimated precisely for the second pulse when the two pulses overlap.
 }


\begin{document}
\maketitle
\flushbottom

\section{Introduction}
Frequency domain multiplexing (FDM) is a technique by which multiple detector signals are combined into a single channel and the combined signal is recorded using a single digitizer input, thereby using fewer digitizer channels for a large number of detector signals. Each individual detector signal can be recovered from the combined signal in the frequency domain, along with the detector number that produced it.

FDM of radiation detectors has been previously implemented for transition-edge sensor (TES) calorimeters by modulating the carrier current flowing through each TES in the time domain\cite{Lanting2005,DenHartog2012,VanDerKuur2009}. The carrier current is a sinusoid of a   specific frequency ($f_c$) that gives a 'tag' unique to each sensor, and the amplitude of the carrier is modulated by the sensor signal $x(t)$ to give an output $y(t)$:
\begin{linenomath*}
\begin{equation}
  y(t) = x(t)\sin(2\pi f_ct)
\end{equation}
\end{linenomath*}
The multiplication of two signals in the time domain is equivalent to their convolution in the frequency domain \cite{dsp}:
\begin{linenomath*}
\begin{equation}
  \begin{split}
  Y(f) = \mathcal{F}(y(t)) & = \int_{-\infty}^{+\infty}y(t)e^{-i2\pi ft}dt \\ & = \frac{1}{2i}\int_{\lambda=-\infty}^{+\infty} \left(\delta(\lambda - f_c) - \delta(\lambda + f_c)\right)X(f - \lambda)d\lambda \\ 
  & = \frac{X(f-f_c) - X(f+f_c)}{2i}\label{ee1}
  \end{split}
\end{equation}
\end{linenomath*}
where $\delta$ is the Dirac delta function, and $\mathcal{F}$ is the Fourier transform operator. As a result, modulation of the sinusoidal carrier signal by its sensor signal moves the spectrum of the sensor signal from its baseband to a passband centered at frequency $f_c$ of the carrier, shown by Eq. \ref{ee1}. All the amplitude-modulated carrier currents flowing through their respective TESs are summed into a single channel and finally demodulated to recover each individual TES signal \cite{Smecher2012}. In order to recover the sensor signal from the modulated signal, the positive frequency components of $Y(f)$ are shifted back to the baseband and the inverse Fourier transform\footnote{$\mathcal{F}^{-1} = \int_{-\infty}^{+\infty}e^{i2\pi ft}df$ is the inverse Fourier transform operator.} is applied:
\begin{linenomath*}
\begin{equation}
  x(t) = 2i\mathcal{F}^{-1}(Y(f+f_c)u(f+f_c))
\end{equation}
\end{linenomath*}
where $u(f)$ is the unit step function.

Amplitude modulation cannot be implemented for pulse mode detectors, so we recently demonstrated a new method of frequency domain multiplexing using convolution and deconvolution \cite{MISHRA201957,Mishra2018}. Each detector was connected to a resonator circuit whose impulse response $h_r$ was a damped sinusoid of a unique frequency. The detector pulses acted as input to the resonator whose output was the convolution between the pulse and the resonator's impulse response:
\begin{linenomath*}
\begin{equation}
  y(t) = x(t)*h_r(t) = \int_{\tau=0}^{t} x(\tau)h_{r}(t - \tau) 
\end{equation}
\end{linenomath*}
The equivalent output $Y(f)$ in the frequency domain is the product of the impulse response $H_r(f)$ and the input $X(f)$:
\begin{linenomath*}
\begin{equation}
  Y(f) = H_{r}(f)X(f)
\end{equation}
\end{linenomath*}
The detectors to be multiplexed were connected to resonators with different oscillation frequencies. The frequency of the resonator output assigned a unique label to each detector. When one of the multiplexed detectors produced a signal, the resulting damped sinusoidal output was combined into a single digitizer channel by a fan-in circuit. 

\begin{figure}[H]
 \includegraphics[width=0.8\linewidth]{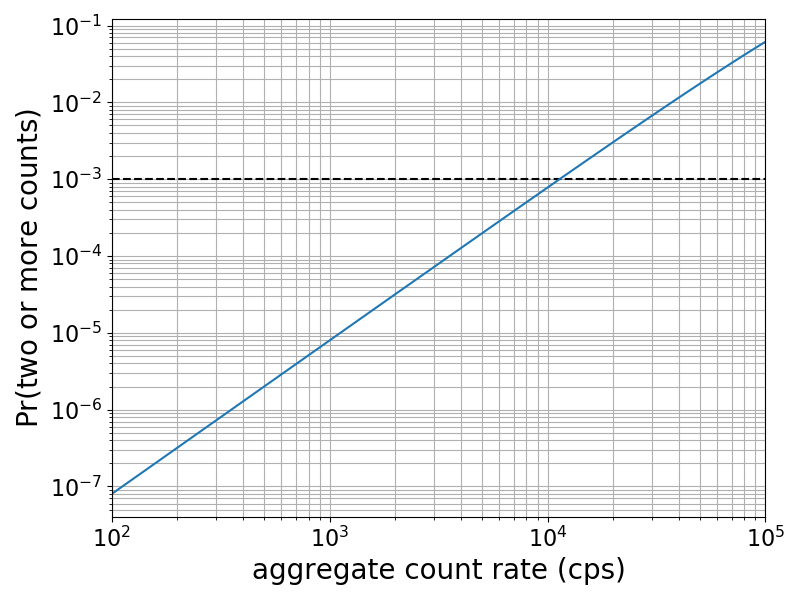}
 \centering
 \caption{The Poisson probability of more than one count in a single digitizer record length of 4 $\mu$s. This probability is 0.001 for count rates $\approx$ 12000 cps.}
 \label{fig:multhits}
\end{figure}

The original detector pulse was then recovered back from the damped sinusoidal output by deconvolution:
\begin{linenomath*}
\begin{equation}
  X(f) = Y(f)/H_{r}(f) 
\end{equation}
\end{linenomath*}
The recovered signal $x(t) = \mathcal{F}^{-1}(X(f))$ was used to estimate the charge collected, time-of-arrival and particle identification of the original pulse. This deconvolution method was meant for the case where a single event occurs in any digitized record. The multiplexer was designed to acquire data for radiation detection experiments with $\lesssim$ $10^{4}$ counts per second (cps); it can be shown that the Poisson probability of the occurrence of two or more events producing signals in a single digitizer record (of length 4 $\mu$s) for count rates less than 12,000 cps is less than one in a thousand (Fig. \ref{fig:multhits}).

In this paper, a new variant of the deconvolution method is demonstrated to enable the recovery of the first signal when multiple detectors produce signals in the same digitized record. The pulse from the first detector is recovered from the digitizer record, and the recovered pulse is used to estimate the charge collected, timing and particle identification of the original pulse. When the pulses produced by two detectors overlap in time, the first pulse can still be partially recovered along with the detector number.

\section{Other multiplexing schemes for pulse mode radiation detector}

The schemes currently used for the multiplexing of pulse mode radiation detectors mostly employ Anger logic/resistive charge-division for silicon photomultiplier (SiPM)-based time-of-flight positron emission tomography (TOF-PET) systems to estimate the interaction location and energy; the timing is encoded on the leading edge of another digital pulse using a high-speed comparator \cite{article_hmSiPM, SiPM_dual}. To minimize the effect of dark noise, only a few SiPM channels are combined to a single comparator and then the comparator outputs are combined into a single channel; the charge is encoded into a digital pulse via time-over-threshold (ToT) \cite{cates}. In this case, a coincidence timing resolution (CTR) of 71 ps at 511 keV was achieved for 2 to 1 multiplexing of LGSO (each coupled to a SensL MicroFC-30035 SiPM). The minimum time delay between two successive pulses should be four times the width\footnote{The pulse width is defined as the duration from the leading edge of the pulse to the time it reaches 0.1 \% of its peak amplitude on the tail end.} of the scintillator pulse for these TOF-PET systems. Capacitive charge-division circuits, which reduce the accumulation of dark current as well as the summed capacitive load, have also been used \cite{choe}. Pulse-tagging multiplexing adds a tag signal of a unique width and height ahead of the scintillation signal to identify the location of firing SiPM; a CRT of $\sim$99 ps at 511 keV was achieved for 2 to 1 multiplexing of LGSO (each coupled to SiPM)\cite{ko_tl}. Signals can also be tagged in frequency by mixing silicon photomultiplier pulse with a sinusoid of specific frequency\cite{wonders}.
The methods discussed above are designed for single-event multiplexing. When pulses overlap, it becomes difficult to estimate the interaction location, so the recovery of energy and timing of the overlapping pulses are not applied using these techniques.

\section{FDM by convolution and deconvolution under multiple occupancy}
In this section, we discuss pulse recovery when two detectors produce signals in the same digitized record, which we refer to as multiple occupancy, for the case when the signals do not overlap in time. 

\begin{figure}[H]
  \includegraphics[width=1.0\linewidth]{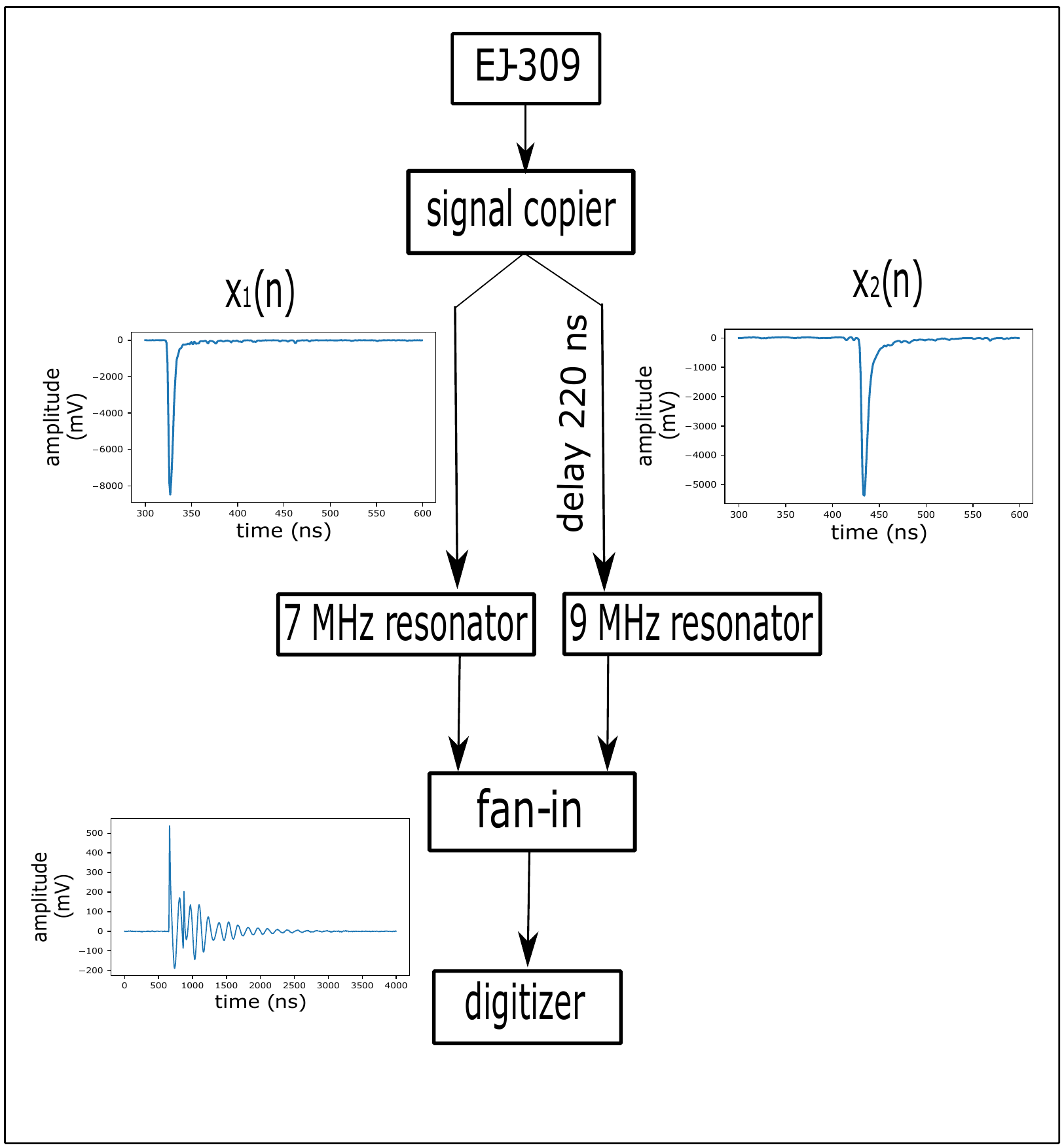}
  \caption{The double occupancy setup. An EJ-309 detector is connected to the signal copier circuit that produces two copies of a detector pulse. The first copy is input to the 7 MHz resonator, and the second copy (after being delayed by 220 ns) is input to the 9 MHz resonator. The resonator outputs are then combined by the fan-in circuit and the fan-in output goes to a single digitizer input channel.}
  \label{fig:p3}
\end{figure}

\begin{figure}[H]
  \includegraphics[width=0.8\linewidth]{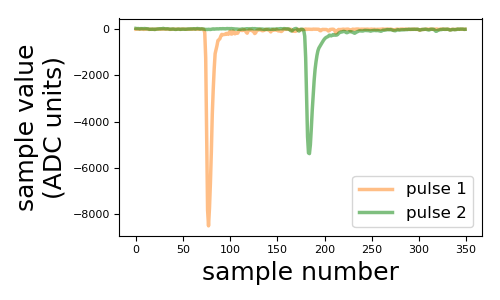}
  \caption{The original pulses $x_1(n)$ and $x_2(n)$ shown together. The second pulse arrives after the first pulse has decayed to zero. The second pulse $x_2(n)$ undergoes $\sim$ 35\% attenuation and bandwidth reduction after passing through the passive delay box.}
  \label{fig:ipulses}
\end{figure}

\begin{figure}[H]
  \centering
  \begin{subfigure}[t]{0.45\linewidth}
    \includegraphics[width=\linewidth]{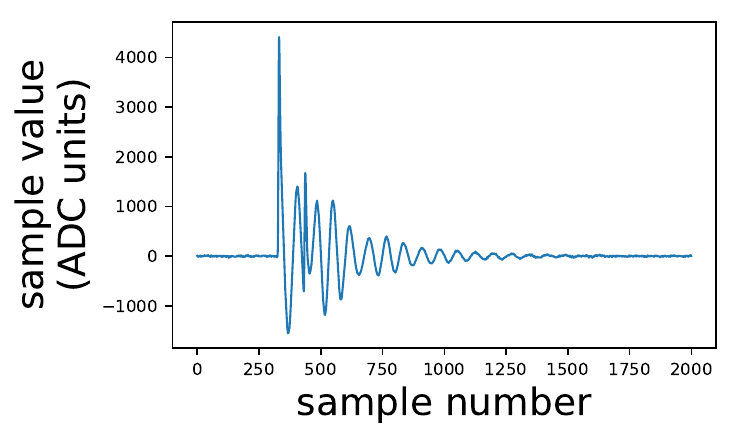}
     
    \caption{The fan-in output when two pulses arrive together in the same digitized record. The damped sinusoids from the 7 MHz and the 9 MHz resonators are combined together.}
    \label{fig:sine1}
  \end{subfigure}
  \hfill
  \begin{subfigure}[t]{0.45\linewidth}
    \includegraphics[width=\linewidth]{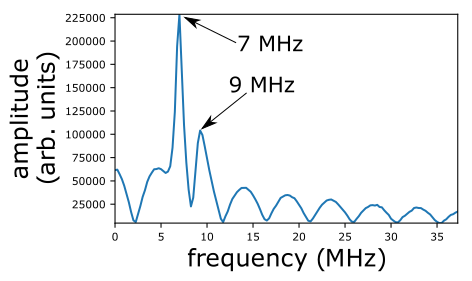}
    
    \caption{The spectrum of the fan-in output shown in Fig. \ref{fig:sine1}. The peaks reveal the resonators that produced the signals, which in turn reveal the detector numbers that produced the signals. In our case,  the signal sources were the two copies of a detector pulse produced by the signal copier circuit. }
    \label{fig:sine2}
  \end{subfigure}
  \caption{The fan-in output in case of double occupancy.}
  \label{fig:sine}
\end{figure}

The setup is shown as a block diagram in Fig. \ref{fig:p3}. A single 7.6 cm x 7.6 cm EJ-309 detector, coupled to an Electron Tube 9821KEB photomultiplier tube (PMT), which produces pulses of about 220 ns wide, with a rise and fall time of about 5 and 20 ns respectively, was used to measure a Cs-137 source. We constructed a `signal copier' circuit to generate two copies of each detector pulse. This circuit accepts a detector pulse as input to two high-speed operational amplifiers in a non-inverting configuration to produce two outputs. Each output signal is a copy of the input signal with the same shape and amplitude. The first copy $x_1(n)$ of the detector pulse was input to a 7 MHz resonator. The second copy $x_2(n)$ was delayed by 220 ns (i.e., the pulse width) using a passive delay line and then input to a 9 MHz resonator. The resonator outputs were combined using the fan-in circuit, and the combined signal was digitized using a CAEN DT5730B, 500 MS/s digitizer with an input range of 500 mVpp.  
Each resonator circuit had a pass-through connector to simultaneously digitize the input pulses along with the fan-in output (see Section 3 for further details of the circuit design); two pulses digitized using their respective resonator pass-throughs are shown in Fig. \ref{fig:ipulses}. The second pulse $x_2(n)$ exhibited $\sim$ 35\% attenuation and bandwidth reduction after passing through the passive delay line, which is clearly visible in Fig. \ref{fig:ipulses}. The digitized fan-in output, which is the sum of two damped sinusoids of frequencies 7 and 9 MHz is shown in Fig. \ref{fig:sine1}. The corresponding spectrum shown in Fig. \ref{fig:sine2} has two peaks at 7 and 9 MHz respectively. 

\subsection{Sequential deconvolution method}
The fan-in output in the discrete-time domain in case of double occupancy is given by
\begin{linenomath*}
\begin{equation}
  y(n) = \sum_{m=0}^{n} x_1(m)h_{r1}(n - m) + \sum_{m=0}^{n} x_2(m)h_{r2}(n - m)\qquad n=0,1,...,N-1
\end{equation}
\end{linenomath*}
where $h_{r1}$, $h_{r2}$ are the impulse responses of the 7 MHz and the 9 MHz resonators, $x_1(n)$, $x_2(n)$ are the inputs to the respective resonators, and $N$ = 2000 is the record length of the digitizer. The discrete Fourier transform of $y(n)$ is given by
\begin{linenomath*}
\begin{equation}
  Y(k) = \sum_{n=0}^{N-1} y(n)exp(\frac{-j2\pi kn}{N}) \qquad k=0,1,...,N-1
\end{equation}
\end{linenomath*}
In the frequency domain, the fan-in output $Y(k)$ is equivalent to the sum of the products of $H_{r1}(k)$ with $X_1(k)$ and $H_{r2}(k)$ with $X_2(k)$:
\begin{linenomath*}
\begin{equation}
  Y(k) = H_{r1}(k)X_1(k) + H_{r2}(k)X_2(k) \qquad k=0,1,...,N-1
\end{equation}
\end{linenomath*}

Both signals can be recovered because $x_1(n)$ and $x_2(n)$ are separable in time. Since $x_1(n)$ arrives first in time, deconvolution with respect to 7 MHz resonator gives 
\begin{linenomath*}
\begin{equation}
  Y_1(k) = \frac{Y(k)}{H_{r1}(k)} = X_1(k) +  \frac{H_{r2}(k)X_2(k)}{H_{r1}(k)} \label{y1}
\end{equation}
\end{linenomath*}
Since $x_1(n)$ and $x_2(n)$ are separable in time, the two terms on the right hand side of Eq. \ref{y1} are also separable in the time domain. Therefore, the minimum time delay between the two signals should be equal to the pulse width of the scintillator, which in this case is about 220 ns. Taking the inverse Fourier transform of Eq. \ref{y1}:
\begin{linenomath*}
\begin{equation}
\begin{split}
  y_1(n) & = \mathcal{F}^{-1}\left(X_1(k)\right) + \mathcal{F}^{-1}\left( \frac{H_{r2}(k)X_2(k)}{H_{r1}(k)}\right) \\                                      & = x_1(n) + \mathcal{F}^{-1}\left( \frac{H_{r2}(k)X_2(k)}{H_{r1}(k)}\right)            \qquad n=0,1,...,N-1 \label{y2}
  \end{split}
\end{equation}
\end{linenomath*}

\begin{figure}[H]
  \centering
  \begin{subfigure}[t]{0.75\linewidth}
    \includegraphics[width=\linewidth]{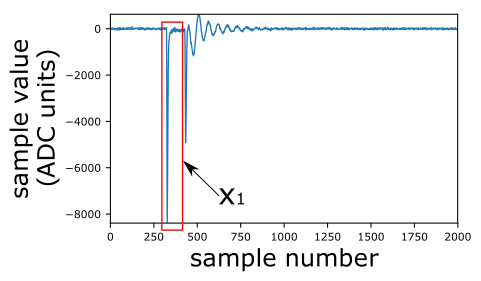}
     
    \caption{The deconvolved signal $y_1(n)$ shows the pulse $x_1(n)$ between sample numbers 300 and 430, followed by the second term $\mathcal{F}^{-1}\left( \frac{H_{r2}(k)X_2(k)}{H_{r1}(k)}\right)$.}
    \label{fig:dsine1}
  \end{subfigure}
  \hfill
  \begin{subfigure}[t]{0.75\linewidth}
    \includegraphics[width=\linewidth]{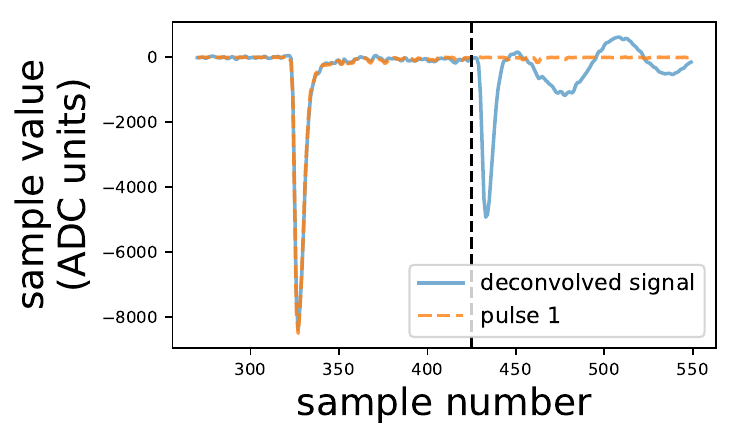}
    
    \caption{The signal $y_1(n)$ shown around the region where the pulse $x_1(n)$ is clearly visible. $x_1(n)$ is recovered from this region between sample numbers 300 and 430. The black dotted line shows sample number where the second pulse $x_2(n)$ arrives. The sinusoidal signal after the black dotted line is the second term $\mathcal{F}^{-1}\left( \frac{H_{r2}(k)X_2(k)}{H_{r1}(k)}\right)$. The original pulse $x_1(n)$ is also shown in orange for reference.}
    \label{fig:dsine2}
  \end{subfigure}
  \caption{The deconvolved signal $y_1(n)$.}
  \label{fig:dsine}
\end{figure}

\begin{figure}[H]
  \includegraphics[width=0.8\linewidth]{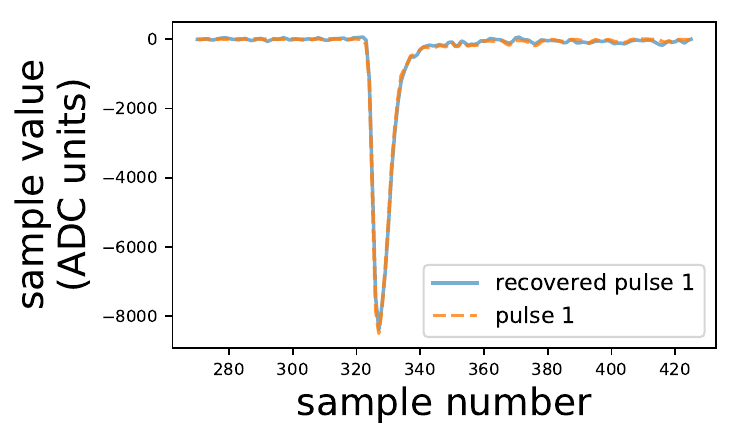}
  \caption{Comparison of the original pulse to the recovered pulse $x_1(n)$.}
  \label{fig:recovers1}
\end{figure}

Fig. \ref{fig:dsine1} shows the deconvolved signal $y_1(n)$ in the time domain. The pulse $x_1(n)$ lies between sample numbers 300 and 430, and the second term $\mathcal{F}^{-1}\left( \frac{H_{r2}(k)X_2(k)}{H_{r1}(k)}\right)$ appears later as seen in Fig. \ref{fig:dsine}. The second term is a function of the pulse $X_2(k)$ in the frequency domain, which appears in $y_1(n)$ when the second pulse $x_2(n)$ arrives. The pulse $x_1(n)$ is recovered from $y_1(n)$ as shown in Fig. \ref{fig:dsine2}. Fig. \ref{fig:recovers1} shows the comparison between the original and the recovered pulse $x_1(n)$.

\begin{figure}[H]
  \includegraphics[width=0.8\linewidth]{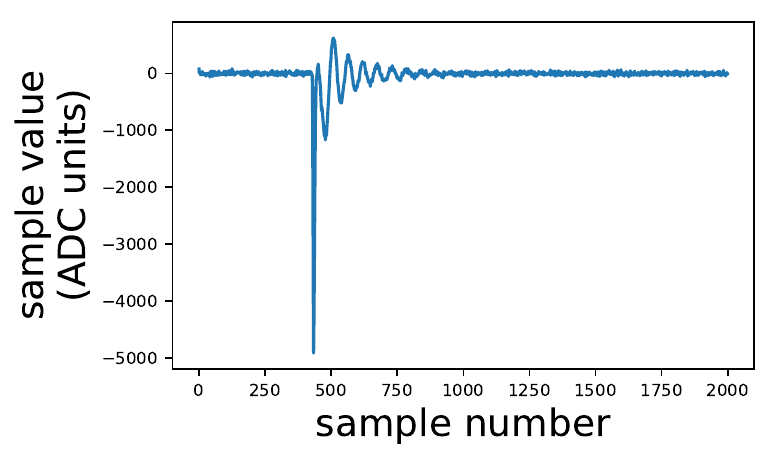}
  \caption{The subtracted deconvolved signal $y_1(n)-x_1(n)$. In the frequency domain, it is equivalent to $Y_1(k) - X_1(k)$ given by Eq. \ref{y33}.}
  \label{fig:subdec_sine1}
\end{figure}

The recovered signal $x_1(n)$ is then subtracted from the deconvolved signal in the time domain and the resulting signal is shown in Fig. \ref{fig:subdec_sine1}. The Fourier transform is a linear operator, so the subtraction of $x_1(n)$ from $y_1(n)$ is equivalent to the subtraction of $X_1(k)$ from $Y_1(k)$. Converting $y_1(n)-x_1(n)$ in Fig. \ref{fig:subdec_sine1} to the frequency domain and using Eq. \ref{y1}: 
\begin{linenomath*}
\begin{equation}
   Y_1(k) - X_1(k) = \frac{Y(k)}{H_{r1}(k)} - X_1(k) = \frac{H_{r2}(k)X_2(k)}{H_{r1}} \label{y33}
\end{equation}
\end{linenomath*}

\begin{figure}[H]
  \includegraphics[width=0.8\linewidth]{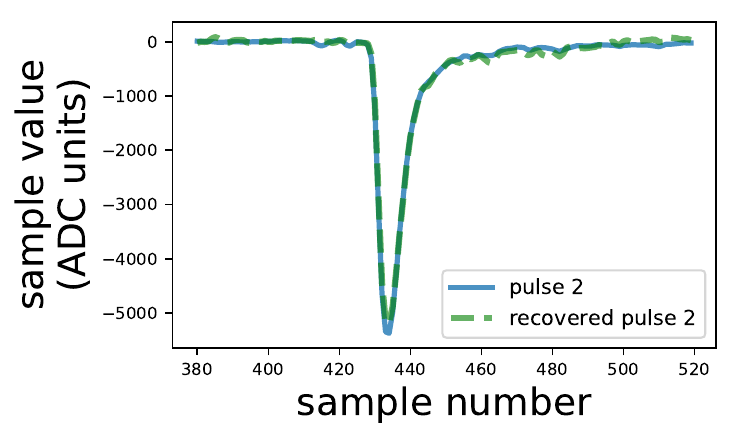}
  \caption{Comparison of the original pulse to the recovered pulse $x_2(n)$.}
  \label{fig:recovers2}
\end{figure}

To recover $x_2(n)$, the terms in Eq.\ref{y33} are rearranged shown below:
\begin{linenomath*}
\begin{equation}
   X_2(k) = \frac{H_{r1}(k)}{H_{r2}(k)}\left(Y_1(k) - X_1(k)\right)  \label{y4}
\end{equation}
\end{linenomath*}
The discrete-time detector signal $x_2(n) =  \mathcal{F}^{-1}\left(X_2(k)\right)$ can finally be recovered by performing the inverse Fourier transform.  Fig. \ref{fig:recovers2} shows the comparison between the original and the recovered pulse $x_2(n)$.

\section{Circuit design}
\begin{figure}[H]
  \centering
  \begin{subfigure}[t]{0.45\linewidth}
    \includegraphics[width=\linewidth]{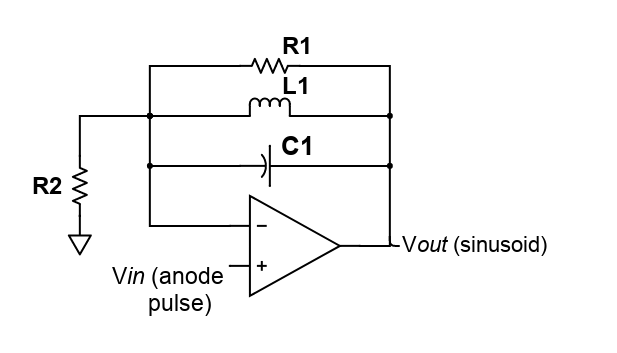}
     
    \caption{The resonator circuit, a parallel RLC circuit designed with a high-speed operational amplifier.}
    \label{fig:osc1}
  \end{subfigure}
  \hfill
  \begin{subfigure}[t]{0.45\linewidth}
    \includegraphics[width=\linewidth]{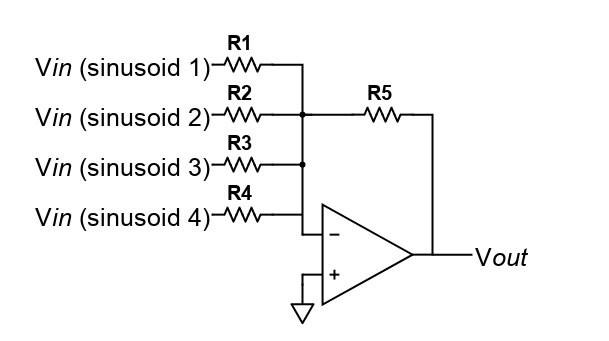}
    
    \caption{The fan-in circuit is a summing amplifier with a gain of 2 ($R_1 = R_2 = R_3 = R_4 = R_5/2$).}
    \label{fig:fan1}
  \end{subfigure}
  \caption{The schematic diagram of the circuits.}
  \label{fig:schematic}
\end{figure}

We designed the resonators and the fan-in circuit (Fig. \ref{fig:schematic}) on a two-sided printed circuit board as explained in \cite{MISHRA201957}. We chose the decay time of the damped sinusoid $<$ 1.5 $\mu$s for each resonator, which resulted in $R_1C_1$ $<$ 0.233 $\mu$s so that the sinusoids completely decay to zero within the digitizer record. This ensured that the resonator output is completely acquired in a digitized record to be able to perform deconvolution. We chose resonant frequencies high enough (starting from 7 MHz) so that the Q-factor = $R\sqrt{\frac{C}{L}}$ $>$ 10 for each resonator, which led to the bandwidth $BW\approx\frac{1}{2\pi Q\sqrt{LC}}\ll 2 MHz$. With the bandwidth of the damped sinusoid for each resonator much less than 2 MHz, we chose resonant frequencies of the resonators 2 MHz apart from each other (i.e., 7 MHz, 9 MHz, 11 MHz ..). This ensured that when the resonators of adjacent resonant frequencies produce signals together in the same digitized record (see Fig. \ref{fig:sine1}), the two peaks in the power spectrum are distinct. The identification of the two peaks for the limiting case when two resonators of adjacent resonant frequencies receive signals from the two detectors connected to them is shown in Fig. \ref{fig:sine2}. 

\begin{figure}[H]
  \includegraphics[width=0.7\linewidth]{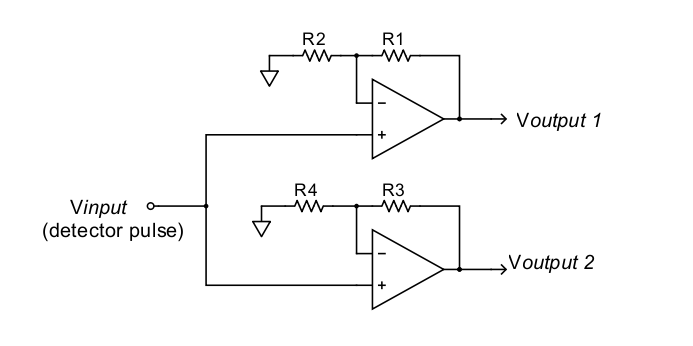}
  \caption{The signal copier circuit ($R_1$ = $R_2$ = $R_3$ = $R_4$).}
  \label{fig:sigcopy}
\end{figure}

The signal copier circuit splits the detector signal to generate two copies of the detector pulse using two high-speed operational amplifiers in a non-inverting configuration with a gain of 2 (Fig. \ref{fig:sigcopy}). 

\section{Impulse response estimation}
To estimate the impulse response of a resonator, each resonator was connected to a fan-in circuit separately. The fan-in output $y(n)$ is the convolution between the input $x(n)$ and the impulse response of the resonator:
\small
\begin{linenomath*}
\begin{equation}
    y(n)=x(n)*h_r(n), Y(k)=X(k)H_r(k) \quad n=0, 1,....N-1 \quad k = 0, 1,....N-1
\end{equation}
\end{linenomath*}
\normalsize
where $N$ = 2000. The input-output cross-correlation function is given by,
\small
\begin{linenomath*}
\begin{equation}\label{eq:crco}
    r_{yx}(m)=r_{xx}(m)*h_r(m), S_{yx}(k)=S_{xx}(k)H_r(k) \quad m=0, 1,....N-1 \quad k = 0, 1,....N-1
\end{equation}
\end{linenomath*}
\normalsize
With a 1 MHz, 1 Vpp noise signal used as input to the resonator, Eq. \ref{eq:crco} can be rearranged to estimate the impulse response:
\begin{linenomath*}
\begin{equation}
\begin{aligned}
  H_r(k) &= \frac{S_{yx}(k)}{S_{xx}(k)} \quad k = 0, 1,....N-1
\end{aligned}
\end{equation}
\end{linenomath*}

\begin{figure}[H]
  \centering
  \begin{subfigure}[t]{0.6\linewidth}
    \includegraphics[width=\linewidth]{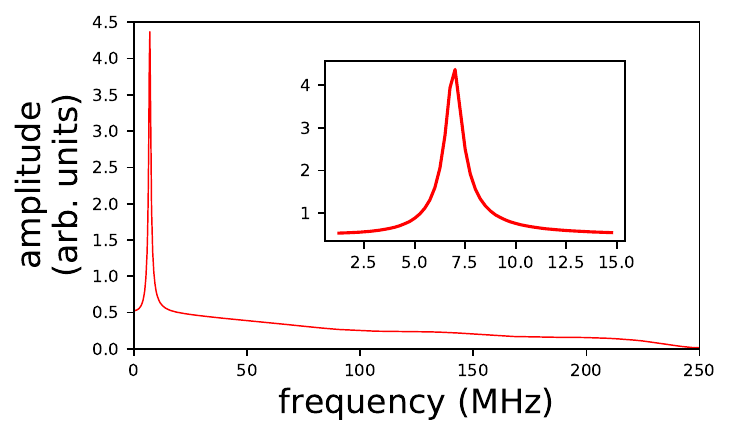}
     
    \caption{The impulse response of the 7.00 MHz resonator in the frequency domain. The inset plot shows the peak at 7 MHz.}
    \label{fig:impres}
  \end{subfigure}
  \hfill
  \begin{subfigure}[t]{0.6\linewidth}
    \includegraphics[width=\linewidth]{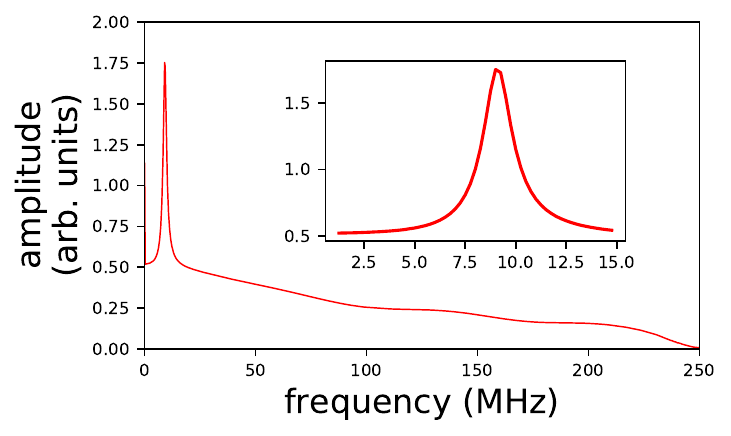}
    
    \caption{The impulse response of the 9 MHz resonator in the frequency domain. The inset plot shows the peak at 9 MHz.}
    \label{fig:impres15}
  \end{subfigure}
  \caption{Resonator impulse response.}
  \label{fig:impress1}
\end{figure}

Fig. \ref{fig:impress1} shows the impulse response of the 7 MHz and 9 MHz resonators. The amplitude of each impulse response above 200 MHz is close to zero because the fan-in output signal becomes indistinguishable from the additive noise of the system. This results in the amplification of the additive noise in the recovered signal X(k), which was not convolved with the impulse response, at these high frequencies after deconvolution \cite{Smith}. This noise is filtered from the recovered signal X(k) by applying an optimized fourth order Butterworth low-pass filter with a cutoff frequency of 160 MHz.

\subsection{Detectability limit on the second pulse amplitude}
\begin{figure}[H]
  \includegraphics[width=0.66\linewidth]{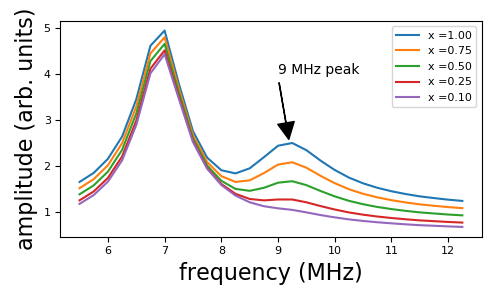}
  \centering
  \captionsetup{justification=centering}
  \caption{The peak at 9 MHz, corresponding to the second pulse, becomes undetectable when the fraction of the second pulse’s
amplitude decreases below 25\% of the first pulse’s amplitude.}
  \label{fig:9mhz_peak}
\end{figure}

\begin{figure}[H]
  \includegraphics[width=0.66\linewidth]{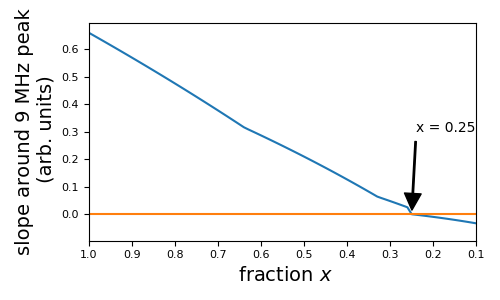}
  \centering
  \captionsetup{justification=centering}
  \caption{ The slope of the 9 MHz peak in the multiplexed signal $Y(k)$ as a function of the amplitude of the second pulse defined by the fraction $x$ of the first pulse.}
  \label{fig:slope}
\end{figure}

Treating the two input pulses as delta functions, with the second pulse amplitude a fraction $x$ of the first one. In the frequency domain, 
\begin{equation}
  Y(k) = H_{r1}(k) + xH_{r2}(k) \qquad k=0,1,...,N-1
\end{equation}
The multiplexed signal $Y(k)$ for different amplitudes of the second pulse defined by the fraction x of the first pulse is shown in the frequency domain in Fig. \ref{fig:9mhz_peak}. As the fraction x decreases from 1, the peak at 9 MHz corresponding to the second pulse decreases in magnitude until the peak vanishes below x = 0.25. The slope of the 9 MHz peak (which is positive) decreases until the peak vanishes at zero when the amplitude of the second pulse is less than 25 $\%$ of the first pulse, after which the second pulse is undetectable (Fig. \ref{fig:slope}). The detectability of the second pulse can be improved when all the channels have equal-magnitude impulse response.

\section{Pulse recovery}
\begin{figure}[H]
  \centering
  \begin{subfigure}[t]{0.65\linewidth}
    \includegraphics[width=\linewidth]{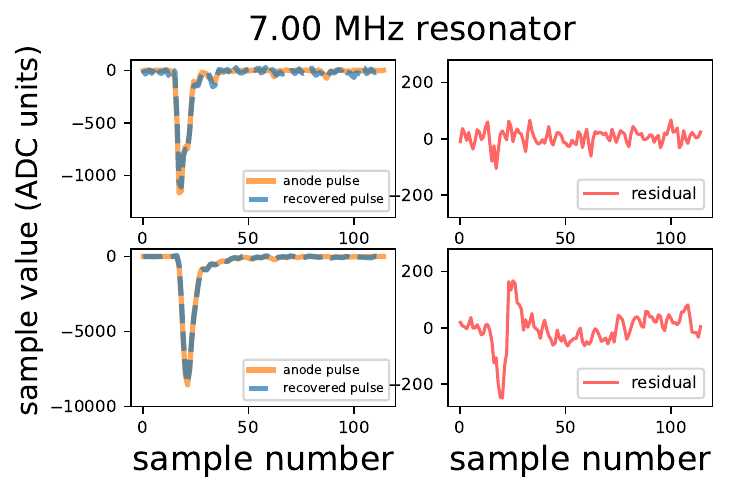}
     
    \caption{The comparison between the anode and the recovered pulse $x_1(n)$ incident on the 7 MHz resonator. The difference between the anode and the recovered pulses is shown on the right.}
    \label{fig:reconst1}
  \end{subfigure}
  \hfill
  \begin{subfigure}[t]{0.65\linewidth}
    \includegraphics[width=\linewidth]{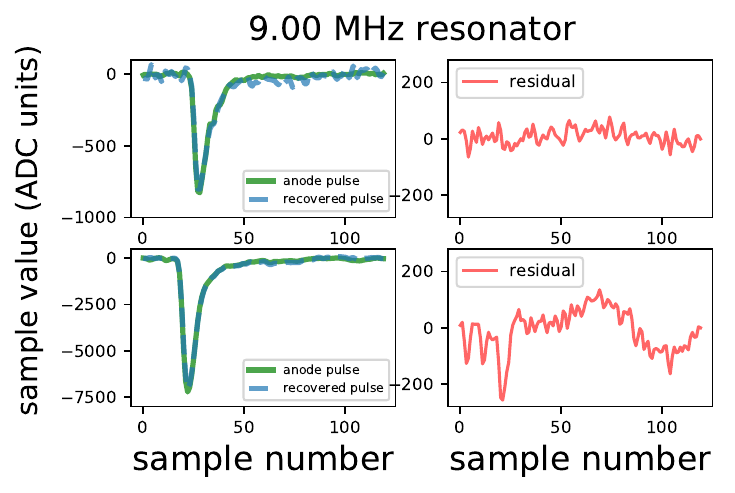}
    
    \caption{The comparison between the anode and the recovered pulse $x_2(n)$ incident on the 9 MHz resonator. The difference between the anode and the recovered pulses is shown on the right.}
    \label{fig:reconst2}
  \end{subfigure}
  \caption{Signal recovery.}
  \label{fig:reconsttt}
\end{figure}

The corresponding recovered signals for pulses $x_1(n)$ and $x_2(n)$ show slight deviation from their original counterparts.The comparison between the original and the recovered pulses is shown in Fig. \ref{fig:reconsttt} for a low amplitude and a high amplitude pulse. The residuals shown for both the pulses exhibit ringing corresponding to the resonant frequency in the additive noise which was not removed by deconvolution. For smaller $x_1(n)$ (around 1500 ADC units), the average rise time and full-width-at-half-maximum for the original and recovered pulses remain the same at 4.97 $\pm$ 0.8 and 10.4 $\pm$ 1.8 ns respectively. For larger pulses (around 7000 ADC units), the average rise time and full-width-at-half-maximum for the original and recovered pulses also remain the same at 4.78 $\pm$ 0.4 and 11.0 $\pm$ 1.7 ns respectively. For smaller $x_2(n)$ (around 1000 ADC units), the average rise time and full-width-at-half-maximum for the original and recovered pulses remain the same at 6.11 $\pm$ 0.8 and 14.0 $\pm$ 1.7 ns respectively. For larger pulses (around 6000 ADC units), the average rise time and full-width-at-half-maximum for the original and recovered pulses also remain the same at 6.02 $\pm$ 0.3 and 14.6 $\pm$ 1.3 ns respectively. 1000 pulses were used to estimate the average rise time and full-width-at-half-maximum for the smaller and larger pulses.

\subsection{Charge estimation using the recovered pulse}
\begin{figure}[H]
  \centering
  \begin{subfigure}[!htb]{0.65\linewidth}
    \includegraphics[width=\linewidth]{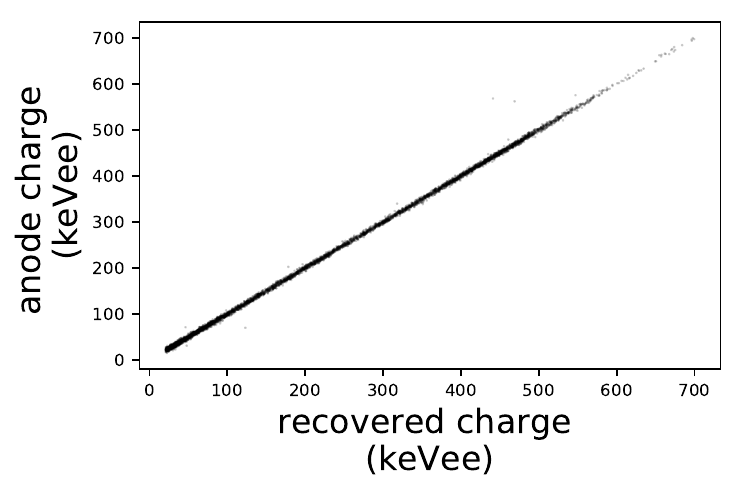}
     
    \caption{Scatter plot between the charge from the first pulse incident on the 7.00 MHz resonator plotted against the recovered charge.}
    \label{fig:ej_ch11}
  \end{subfigure}
  \hfill
  \begin{subfigure}[!htb]{0.65\linewidth}
    \includegraphics[width=\linewidth]{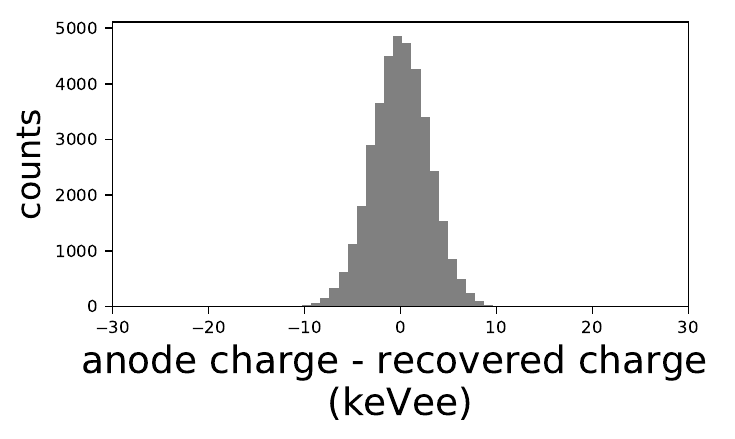}
    
    \caption{The events in Fig. \ref{fig:ej_ch11} shown as the distribution of the difference between the charge from the first pulse and the recovered charge with $\sigma$ =  2.9 $\pm$ 0.02 keVee. A slight bias in the mean is due to the ringing corresponding to the resonant frequency in the additive noise.}
    \label{fig:ej_ch22}
  \end{subfigure}
  \caption{Charge estimation (or area) under the first pulse $x_1(n)$ from the recovered pulse.}
  \label{fig:ej_chh}
\end{figure}

\begin{figure}[H]
  \centering
  \begin{subfigure}[!htb]{0.6\linewidth}
    \includegraphics[width=\linewidth]{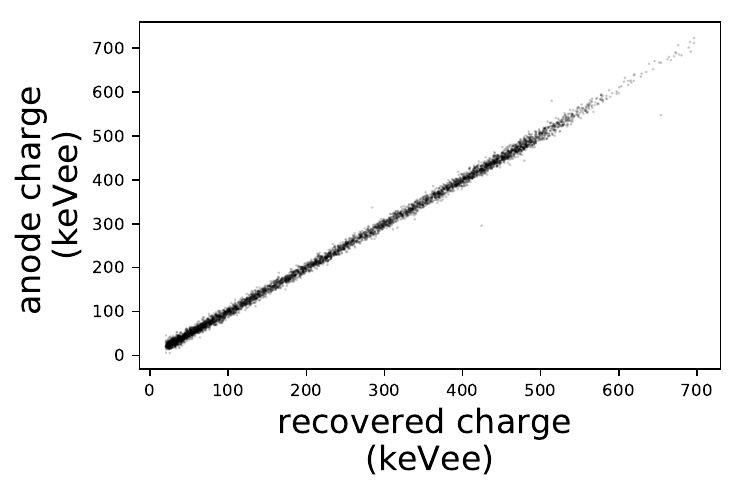}
     
    \caption{Scatter plot between the charge from the second pulse incident on the 9.00 MHz resonator plotted against the recovered charge.}
    \label{fig:ej_ch11p2}
  \end{subfigure}
  \hfill
  \begin{subfigure}[!htb]{0.6\linewidth}
    \includegraphics[width=\linewidth]{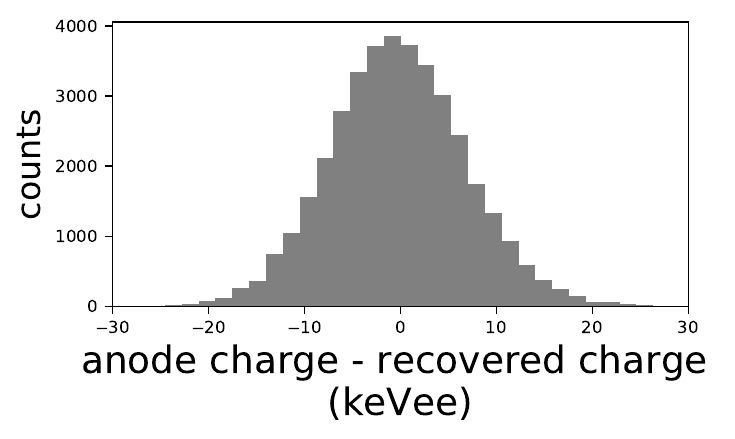}
    
    \caption{The events in Fig. \ref{fig:ej_ch11p2} shown as the distribution of the difference between the charge from the second pulse and the recovered charge with $\sigma$ = 7.1 $\pm$ 0.09 keVee. A slight bias in the mean is due to the ringing corresponding to the resonant frequency in the additive noise.}
    \label{fig:ej_ch22p2}
  \end{subfigure}
  \caption{Charge estimation (or area) under the second pulse $x_2(n)$ from the recovered pulse.}
  \label{fig:ej_chhp2}
\end{figure}

The charge collected under the first pulse $x_1(n)$ can be estimated from the corresponding recovered pulse. Fig. \ref{fig:ej_ch11} shows a scatter plot between the charge under the original first pulse and the charge under the recovered pulse. The uncertainty in the estimate of the charge under the first pulse from the recovered pulse using the histogram in Fig. \ref{fig:ej_ch22} is 2.9 $\pm$ 0.02\footnote{The numbers followed by $\pm$ denote the uncertainty in the uncertainty estimate.} keVee (keV electron-equivalent). Similarly, the uncertainty in the estimate of the charge under the second pulse from the recovered charge using the histogram in Fig. \ref{fig:ej_chhp2} is 7.1 $\pm$ 0.09 keVee, which is nearly twice as large. The increase in uncertainty in the estimate of charge of the second pulse is due to $\sim$ 35\% attenuation and bandwidth reduction encountered by the second pulse after passing through the passive delay line; the amplified noise at high frequencies makes it difficult to precisely calculate the area under the tail of the recovered pulse, particularly because the pulse is attenuated and its tail is elongated.   

\subsection{Time pick-off using the recovered pulse}
\begin{figure}[H]
  \centering
  \begin{subfigure}[!htb]{0.65\linewidth}
    \includegraphics[width=\linewidth]{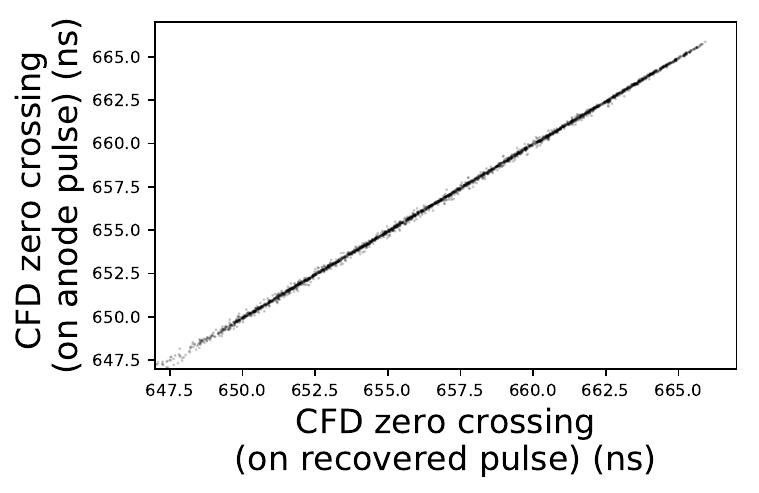}
     
    \caption{Scatter plot between time pick-off of the first pulse incident on the 7.00 MHz resonator plotted against the time pick-off of the recovered pulse.}
    \label{fig:at_rt11}
  \end{subfigure}
  \hfill
  \begin{subfigure}[!htb]{0.65\linewidth}
    \includegraphics[width=\linewidth]{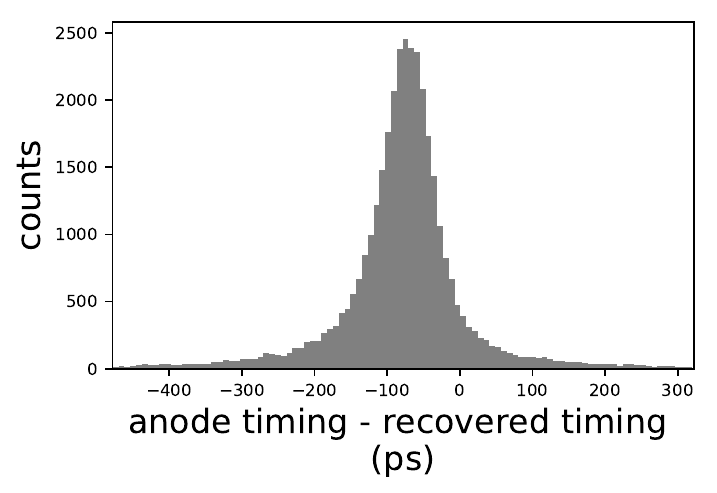}
    
    \caption{The events in Fig. \ref{fig:at_rt11} shown as the distribution of the difference between the timing of the first pulse and the recovered timing with $\sigma$ = 87 $\pm$ 1.3 ps. A small bias in the mean is due to the ringing corresponding to the resonant frequency in the additive noise.}
    \label{fig:at_rt22}
  \end{subfigure}
  \caption{Timing estimation of the first pulse $x_1(n)$ from the recovered pulse.}
  \label{fig:at_rtt}
\end{figure}

Constant fraction discrimination (CFD) was performed on both the pulses and their recovered counterparts to eliminate amplitude-dependent time walk\cite{Fallu-Labruyere2007}. The CFD timing mark was computed by subtracting a delayed copy of the pulse from an attenuated copy and then computing the first zero crossing of the resultant signal. The CFD time pick-off used an attenuation fraction of 0.2 with a 6.4 ns delay for the first pulse. Fig. \ref{fig:at_rt11} shows the scatter plot of the time pick-off between the first pulse and the recovered pulse. The uncertainty in the estimate of the timing of the first pulse from the recovered pulse using the histogram in Fig. \ref{fig:at_rt22} is 87 $\pm$ 1.3 ps. 

\begin{figure}[H]
  \centering
  \begin{subfigure}[!htb]{0.65\linewidth}
    \includegraphics[width=\linewidth]{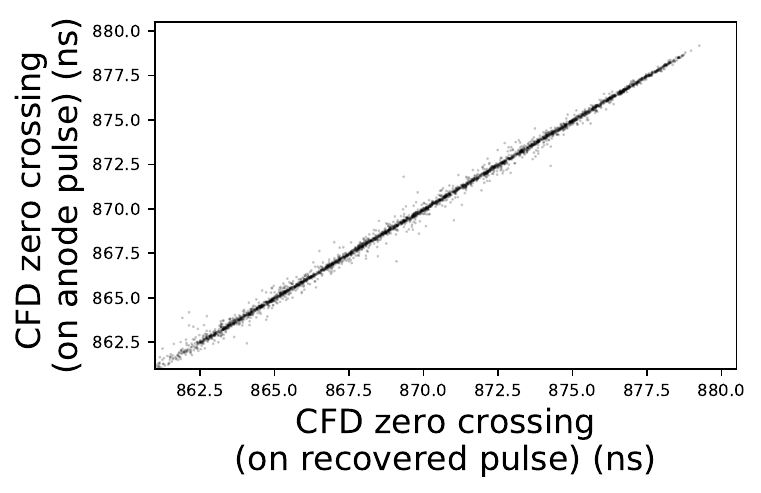}
     
    \caption{Scatter plot between time pick-off of the second pulse incident on the 9.00 MHz resonator plotted against the time pick-off of the recovered pulse.}
    \label{fig:at_rt11p2}
  \end{subfigure}
  \hfill
  \begin{subfigure}[!htb]{0.65\linewidth}
    \includegraphics[width=\linewidth]{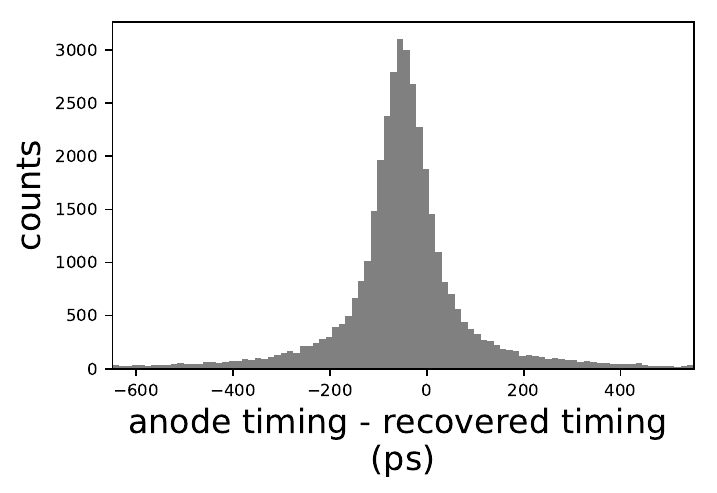}
    
    \caption{The events in Fig. \ref{fig:at_rt11p2} shown as the distribution of the difference between the timing of the second pulse and the recovered timing with $\sigma$ = 137 $\pm$ 4.3 ps. A small bias in the mean is due to the ringing corresponding to the resonant frequency in the additive noise.}
    \label{fig:at_rt22p2}
  \end{subfigure}
  \caption{Timing estimation of the second pulse $x_2(n)$ from the recovered pulse.}
  \label{fig:at_rttp2}
\end{figure}

We used an attenuation fraction of 0.3 with a 6.8 ns delay for the second pulse to account for its attenuation and bandwidth reduction.\footnote{Under normal circumstances, the CFD parameters for both the pulses would be the same; however, we changed the attenuation fraction and the delay for the second pulse to account for the change in the pulse shape after passing through the passive delay line.} The uncertainty in the estimate of the timing of the second pulse from the recovered pulse using the histogram in Fig. \ref{fig:at_rttp2} is 137 $\pm$ 4.3 ps, poorer than for the first pulse.

\subsection{Pulse-shape discrimination (PSD) using the recovered pulse} 
We used a Cf-252 source to induce pulses in an EJ-309 detector using the same setup in Fig. \ref{fig:p3}. The two copies ($x_1(n)$ and $x_2(n)$) of the detector signal generated using the signal copier were recovered from the combined signal using the sequential deconvolution method.

\begin{figure}[H]
  \centering
  \begin{subfigure}[t]{0.75\linewidth}
    \includegraphics[width=\linewidth]{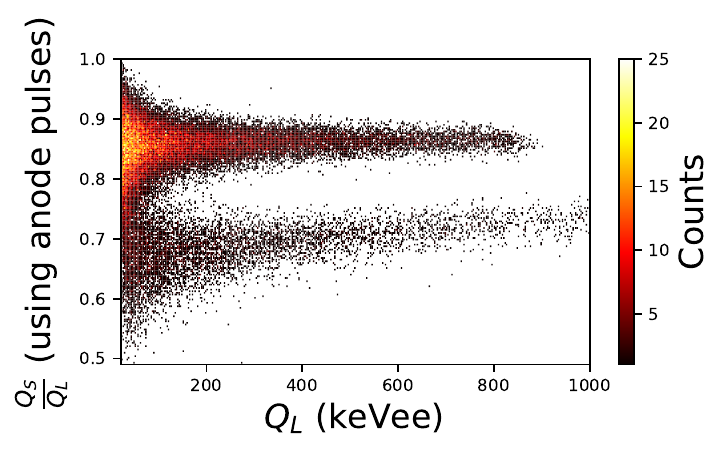}
     
    \caption{Distribution of the discrimination parameter against the total charge $Q_{L}$ using the first pulse.}
    \label{fig:psd11}
  \end{subfigure}
  \hfill
  \begin{subfigure}[t]{0.75\linewidth}
    \includegraphics[width=\linewidth]{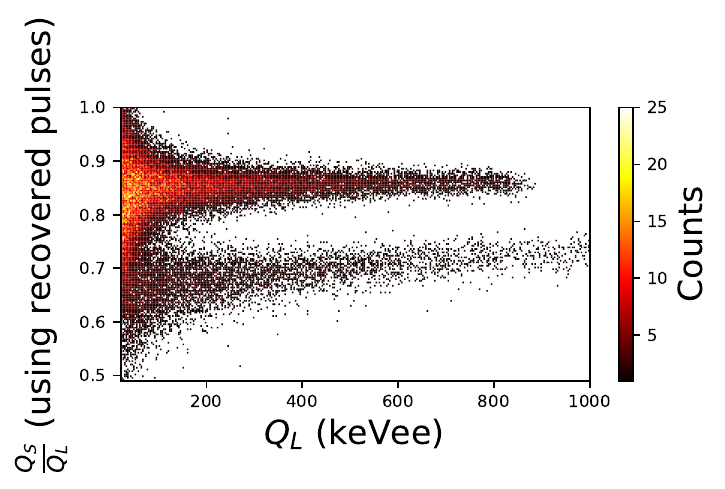}
    
    \caption{Distribution of the discrimination parameter against the total charge $Q_{L}$ using the recovered pulse.}
    \label{fig:psd22}
  \end{subfigure}
  \caption{Pulse shape discrimination for the first pulse.}
  \label{fig:psdsd}
\end{figure}

\begin{figure}[H]
  \centering
  \begin{subfigure}[t]{0.65\linewidth}
    \includegraphics[width=\linewidth]{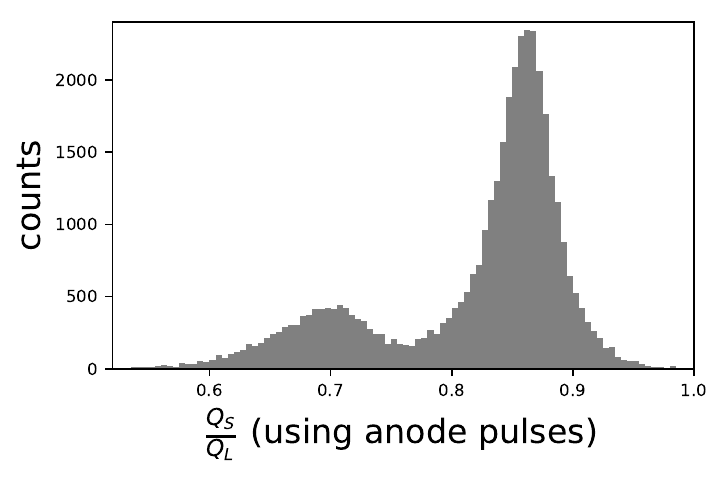}
     
    \caption{Distribution of the discrimination parameter for the first pulse (FOM = 1.01).}
    \label{fig:psdd11}
  \end{subfigure}
  \hfill
  \begin{subfigure}[t]{0.65\linewidth}
    \includegraphics[width=\linewidth]{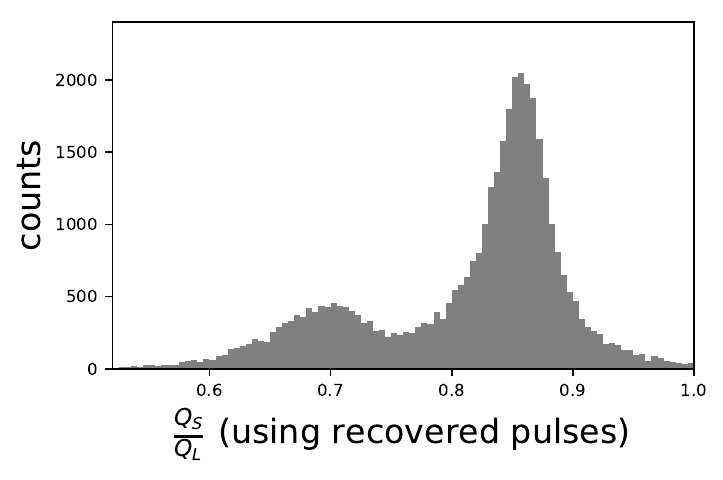}
    
    \caption{Distribution of the discrimination parameter for the recovered pulse (FOM = 0.86).}
    \label{fig:psdd22}
  \end{subfigure}
  \caption{Comparison of the figure of merit for the first pulse incident on the 7 MHz resonator.}
  \label{fig:psddsd}
\end{figure}

\begin{figure}[H]
  \centering
  \begin{subfigure}[t]{0.75\linewidth}
    \includegraphics[width=\linewidth]{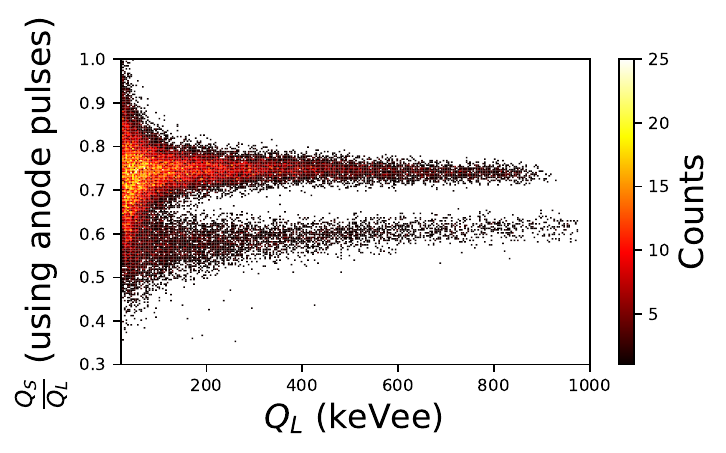}
     
    \caption{Distribution of the discrimination parameter against the total charge $Q_{L}$ using the second pulse.}
    \label{fig:psd111}
  \end{subfigure}
  \hfill
  \begin{subfigure}[t]{0.75\linewidth}
    \includegraphics[width=\linewidth]{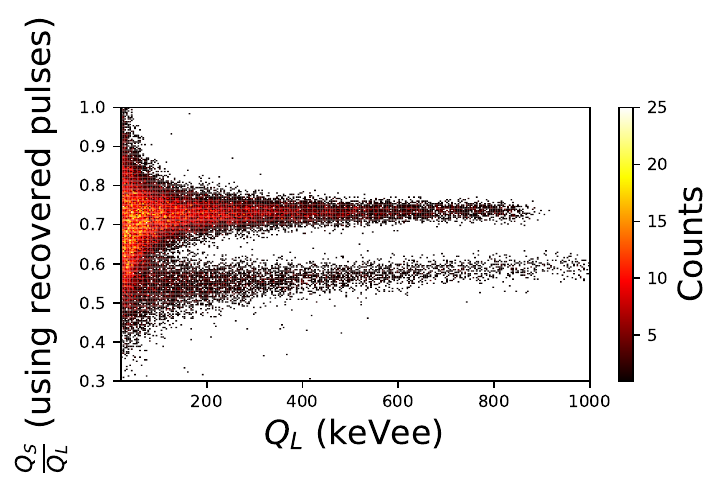}
    
    \caption{Distribution of the discrimination parameter against the total charge $Q_{L}$ using the recovered pulse.}
    \label{fig:psd222}
  \end{subfigure}
  \caption{Pulse shape discrimination for the second pulse.}
  \label{fig:psdsd1}
\end{figure}

\begin{figure}[H]
  \centering
  \begin{subfigure}[t]{0.55\linewidth}
    \includegraphics[width=\linewidth]{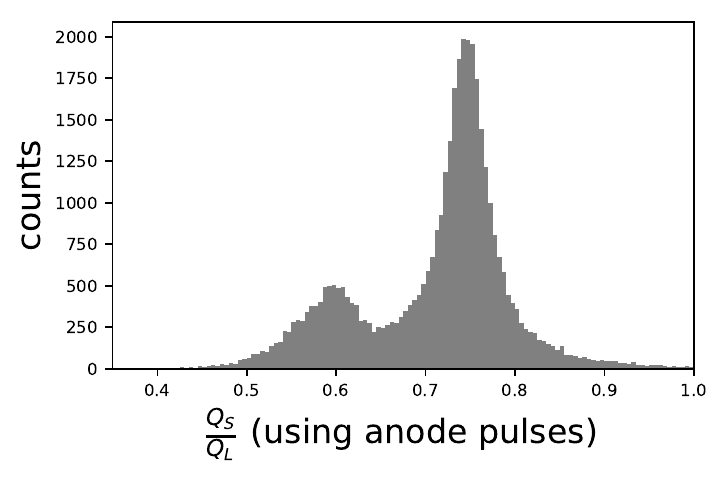}
     
    \caption{Distribution of the discrimination parameter for the second pulse. (FOM = 0.90).}
    \label{fig:psdd111}
  \end{subfigure}
  \hfill
  \begin{subfigure}[t]{0.55\linewidth}
    \includegraphics[width=\linewidth]{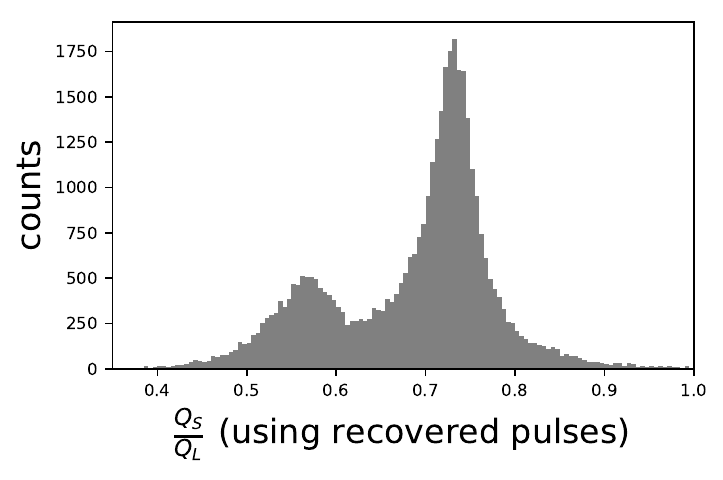}
    
    \caption{Distribution of the discrimination parameter for the recovered pulse (FOM = 0.89).} 
    \label{fig:psdd222}
  \end{subfigure}
  \caption{Comparison of the figure of merit for the second pulse incident on the 9 MHz resonator.}
  \label{fig:psddsd1}
\end{figure}

The pulses generated by neutrons will have a longer tail than the pulses generated by gamma rays, so the charge-integration method was used to discriminate neutron pulses from the gamma pulses. The discrimination parameter was defined as the ratio of the area under the pulse excluding its tail ($Q_{S}$), to the total area ($Q_{L}$) under the pulse. 
We plotted a 2D-histogram of the discrimination parameter $\frac{Q_{S}}{Q_{L}}$ against the total charge (area) $Q_{L}$ under the pulse for the original and the recovered pulses for a threshold of 20 keVee (Fig. \ref{fig:psdsd}, Fig. \ref{fig:psdsd1}). For the first pulse $x_1(n)$, the pulse length was kept constant at 107 samples. The start time of the integration length for $Q_{S}$ was fixed at six samples to the left of the pulse peak, and the stop time  was optimized by selecting the integration length that maximized the figure of merit FOM = ($\mu_{gamma} - \mu_{neutron}$)/($FWHM_{gamma} + FWHM_{neutron}$), where $\mu$ denotes the mean and $FWHM$ denotes the full width at half-maximum\cite{Winyard1971}; the larger the FOM the better the discrimination between the neutron and gamma events. Using the distribution of the discrimination parameter shown in Fig. \ref{fig:psddsd}, an FOM of 1.01 was obtained using the original first pulse incident on the 7 MHz resonator compared to 0.86 when its recovered counterpart was used. For the second pulse $x_2(n)$, the pulse length was kept constant at 177 samples. The start time of the integration length for $Q_{S}$ was fixed at eight samples to the left of the pulse peak, and the stop time was optimized to maximize the FOM. An FOM of 0.90 was obtained using the original second pulse incident on the 9 MHz resonator compared to 0.89 when its recovered counterpart was used (Fig. \ref{fig:psddsd1}). The reduction in FOM for the recovered pulses is due to the ringing corresponding to the resonant frequency in the additive noise (see Fig. \ref{fig:reconsttt}). 
The error in the discrimination parameter $\frac{Q_{S}}{Q_{L}}$ tends to increase with decreasing pulse amplitude. 


\section{Overlapping pulses}\label{overlap}
\begin{figure}[H]
  \includegraphics[width=0.7\linewidth]{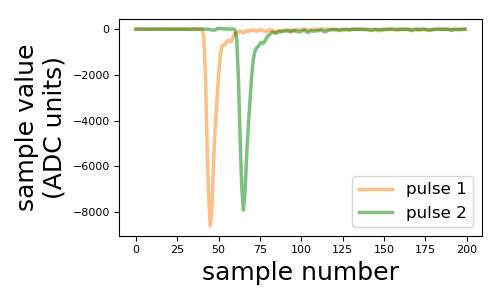}
  \caption{The original pulses $x_1(n)$ and $x_2(n)$ shown together. The second pulse arrives 30 ns after the first pulse.}
  \label{fig:ipulses32}
\end{figure}

\begin{figure}[H]
  \centering
  \begin{subfigure}[t]{0.45\linewidth}
    \includegraphics[width=\linewidth]{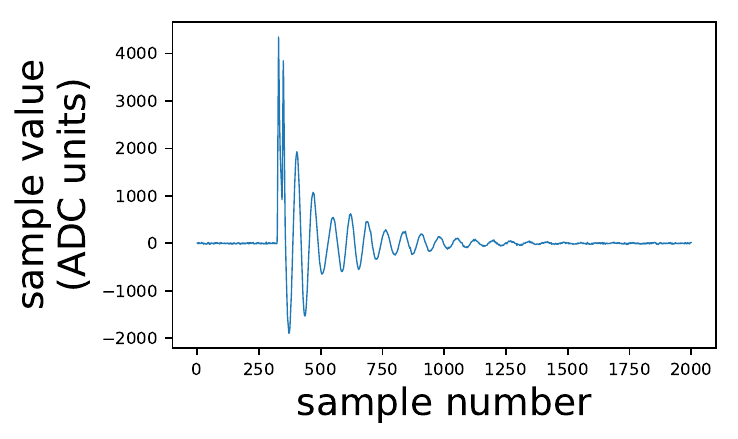}
     
    \caption{The fan-in output when two pulses that overlap in time arrive together in the same digitized record. The damped sinusoids from the 7 MHz and the 9 MHz resonators are combined together.}
    \label{fig:sine132}
  \end{subfigure}
  \hfill
  \begin{subfigure}[t]{0.5\linewidth}
    \includegraphics[width=\linewidth]{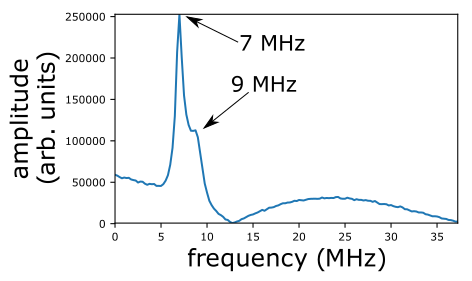}
    
    \caption{The spectrum of the fan-in output shown in Fig. \ref{fig:sine1}. The peaks reveal the resonators that produced the signals, which in turn reveal the detector numbers that produced the signals.}
    \label{fig:sine232}
  \end{subfigure}
  \caption{The fan-in output in case the pulses overlap.}
  \label{fig:sine32}
\end{figure}

We used the same setup shown in Fig. 2 to generate two copies of an EJ-309 pulse using the signal copier circuit. The delay in the second copy $x_2(n)$ was chosen to be 30 ns so that it overlapped with the tail of the first pulse $x_1(n)$ in time as shown in Fig. \ref{fig:ipulses32}. The digitized fan-in output $y(n)$ in this case is shown in Fig. \ref{fig:sine32}.

\begin{figure}[H]
  \centering
  \begin{subfigure}[t]{0.65\linewidth}
    \includegraphics[width=\linewidth]{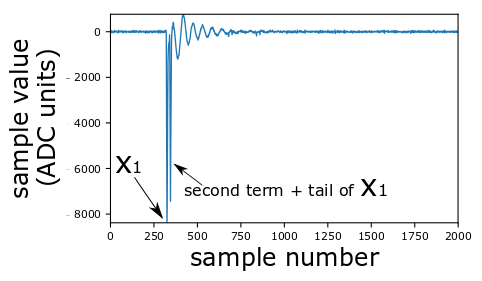}
     
    \caption{The deconvolved signal $y_1(n)$ shows the pulse $x_1(n)$ starting near sample number 320 with the second term $\mathcal{F}^{-1}\left( \frac{H_{r2}(k)X_2(k)}{H_{r1}(k)}\right)$ overlapping its tail.}
    \label{fig:dsine132}
  \end{subfigure}
  \hfill
  \begin{subfigure}[t]{0.65\linewidth}
    \includegraphics[width=\linewidth]{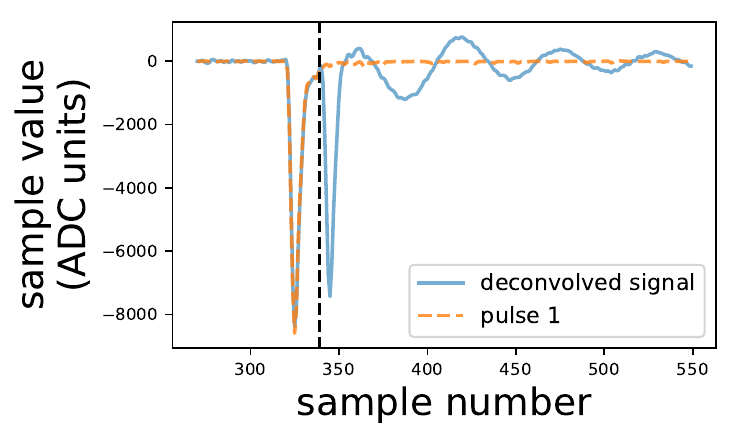}
    
    \caption{The signal $y_1(n)$ shown around the region where the pulse $x_1(n)$ is clearly visible. $x_1(n)$ is recovered from this region from when it arrives near sample number 320 until the arrival of the second pulse 30 ns later. The black dotted line shows sample number from where the second pulse $x_2(n)$ arrives. The sinusoidal signal after the black dotted line is the second term $\mathcal{F}^{-1}\left( \frac{H_{r2}(k)X_2(k)}{H_{r1}(k)}\right)$ contaminated by the additive tail of the  first pulse. The original pulse $x_1(n)$ is also shown in orange for reference.}
    \label{fig:dsine232}
  \end{subfigure}
  \caption{The deconvolved signal $y_1(n)$.}
  \label{fig:dsine32}
\end{figure}

\begin{figure}[H]
  \includegraphics[width=0.8\linewidth]{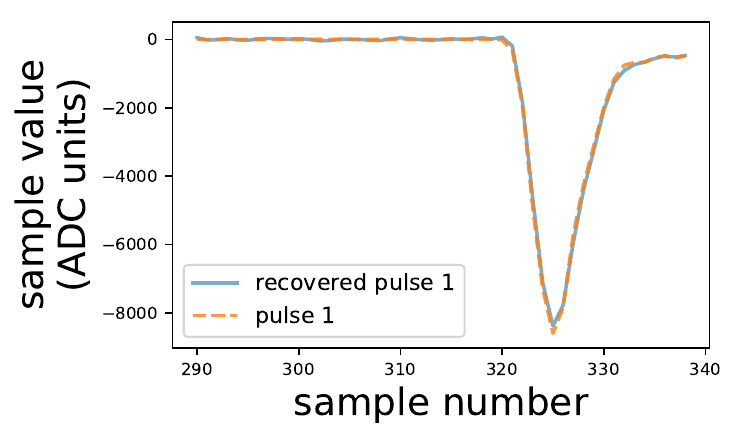}
  \caption{The comparison between the original pulse $x_1(n)$ and the partially recovered pulse $\tilde{x_1}(n)$.}
  \label{fig:recovers132}
\end{figure}

Because $x_1(n)$ and $x_2(n)$ overlap in time, the two terms on the right hand side of Eq. \ref{y1} are not completely separable in the time domain. Fig. \ref{fig:dsine132} shows the deconvolved signal $y_1(n)$ in the time domain. The pulse $x_1(n)$ lying between sample numbers 300 and 430 overlaps the second term of Eq. \ref{y2} $\mathcal{F}^{-1}\left( \frac{H_{r2}(k)X_2(k)}{H_{r1}(k)}\right)$ that arrives on the top of the tail of $x_1(n)$ as shown in Fig. \ref{fig:dsine32}. Therefore, $x_1(n)$ is accurately recoverable only until the arrival of the second pulse. The partial recovery of the pulse $x_1(n)$ from $y_1(n)$ is shown in Fig. \ref{fig:dsine232}. Fig. \ref{fig:recovers132} shows the comparison between the original pulse $x_1(n)$ and the partially recovered pulse $\tilde{x_1}(n)$.

\begin{figure}[H]
  \includegraphics[width=0.8\linewidth]{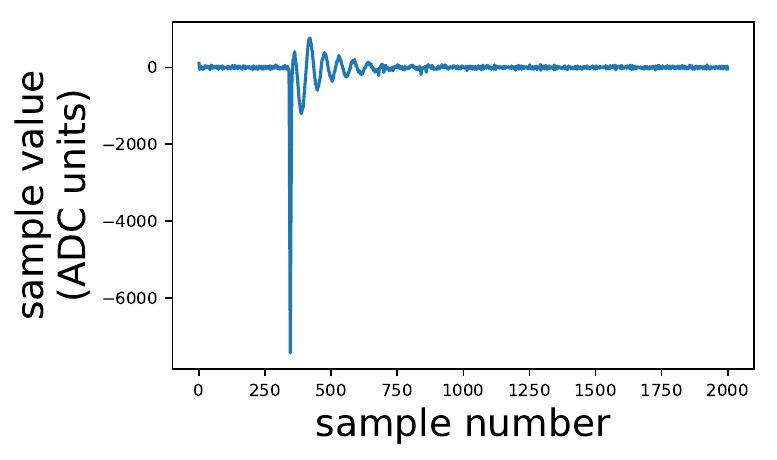}
  \caption{The subtracted deconvolved signal $y_1(n)-\tilde{x_1}(n)$. In the frequency domain, it is equivalent to $Y_1(k) - \tilde{X_1}(k)$ given by Eq. \ref{y3332}.}
  \label{fig:subdec_sine132}
\end{figure}

The partially recovered signal $\tilde{x_1}(n)$ is then subtracted from the deconvolved signal in the time domain and the resulting signal $y_1(n)-\tilde{x_1}(n)$ is shown in Fig. \ref{fig:subdec_sine132}. Converting $y_1(n)-\tilde{x_1}(n)$ in Fig. \ref{fig:subdec_sine132} to the frequency domain and using Eq. \ref{y1}:

\begin{linenomath*}
\begin{equation}
   Y_1(k) - \tilde{X_1}(k) = \frac{Y(k)}{H_{r1}(k)} - \tilde{X_1}(k) =  \frac{H_{r2}(k)X_2(k)}{H_{r1}(k)} + \mathcal{F}\left(\text{tail of } x_1(n)\right) \label{y3332}
\end{equation}
\end{linenomath*}

\begin{figure}[H]
  \includegraphics[width=0.8\linewidth]{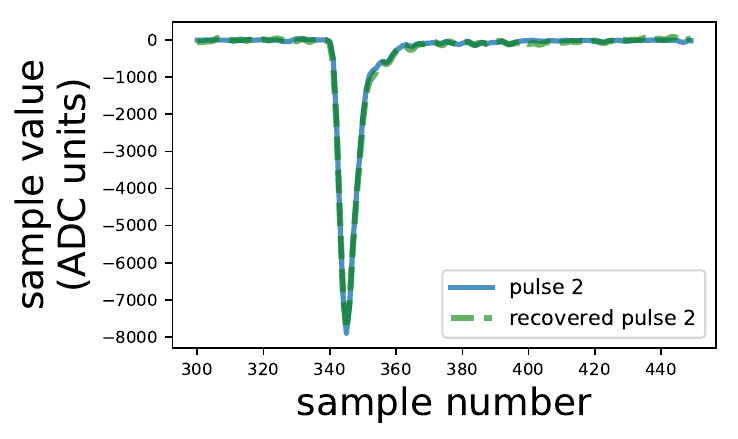}
  \caption{The comparison between the original pulse $x_2(n)$ and the recovered pulse $\tilde{x_2}(n)$.}
  \label{fig:recovers232}
\end{figure}

In this case, when Eq. \ref{y3332} is rearranged as shown below, $x_2(n)$ cannot be accurately recovered  due to the error introduced by the addition of the tail of the first pulse to the second term $\frac{H_{r2}(k)X_2(k)}{H_{r1}(k)}$:
\begin{linenomath*}
\begin{equation}
   \tilde{X_2}(k)=X_2(k)+\frac{H_{r1}(k)}{H_{r2}(k)}\mathcal{F}\left(\text{tail of } x_1(n)\right) = \frac{H_{r1}(k)}{H_{r2}(k)}\left(Y_1(k) - \tilde{X_1}(k)\right)  \label{y432}
\end{equation}
\end{linenomath*}
The error term $\left(\frac{H_{r1}(k)}{H_{r2}(k)}\mathcal{F}\left(\text{tail of } x_1(n)\right)\right)$ increases when $x_1(n)$ and $x_2(n)$ are closer together in time and vanishes when the two pulses cease to overlap in time. The discrete-time detector signal $\tilde{x_2}(n)=x_2(n)+\text{error term = }\mathcal{F}^{-1}\left(\tilde{X_2}(k)\right)$ can be recovered by performing the inverse Fourier transform. Fig. \ref{fig:recovers232} shows the recovered pulse $\tilde{x_2}(n)$ does not accurately follow the original pulse $x_2(n)$, especially on the rising edge. 

\begin{figure}[H]
  \includegraphics[width=0.65\linewidth]{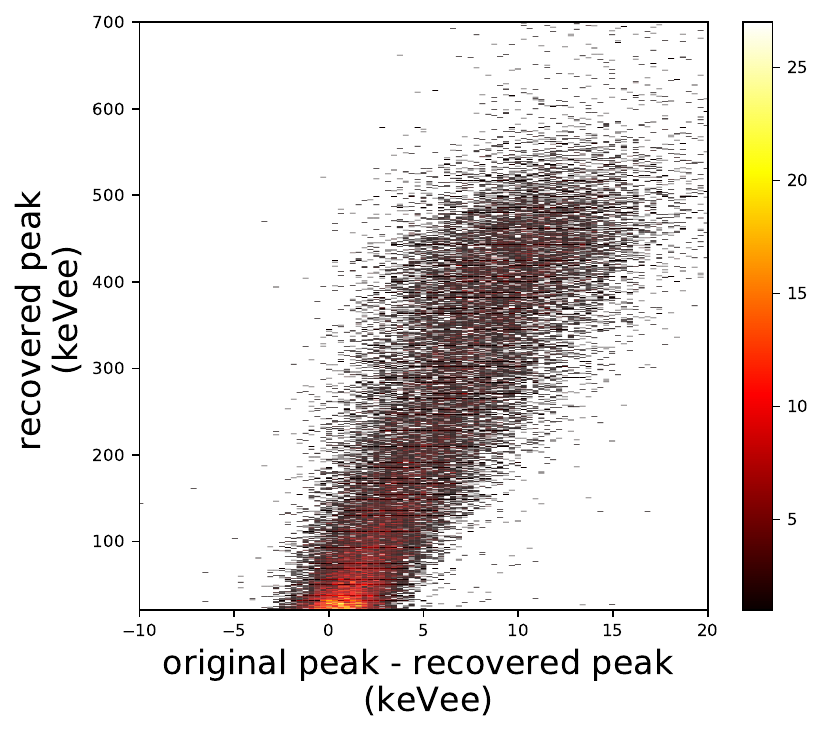}
  \caption{The recovered energy plotted against the difference between the original and recovered energy for the first pulse $x_1(n)$.} 
  \label{fig:ch50_2d}
\end{figure}

\begin{figure}[H]
    \centering 
\begin{subfigure}{0.48\textwidth}
  \includegraphics[width=\linewidth]{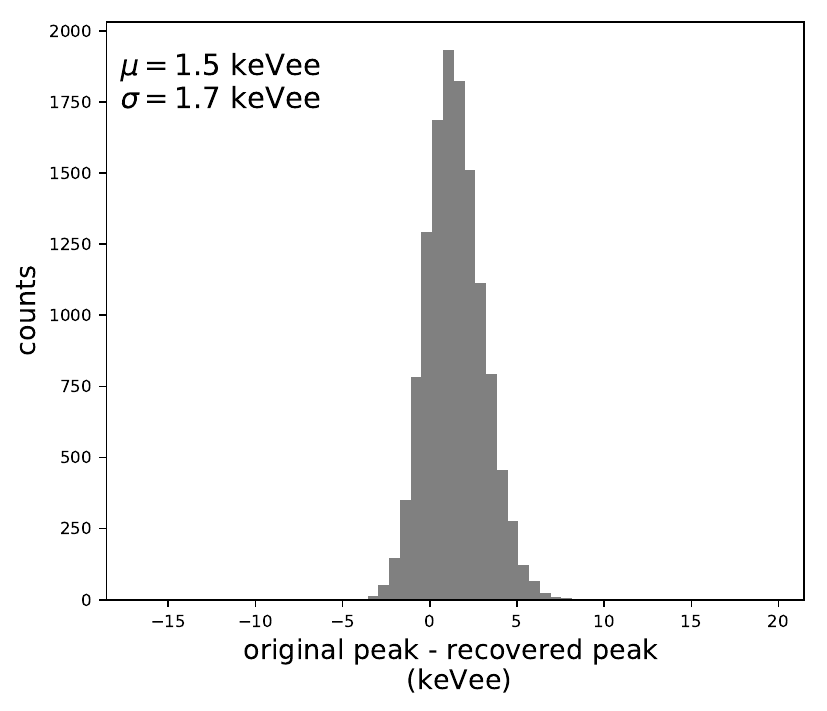}
  \caption{Energy range: 20 - 120 keV}
  \label{fig:150}
\end{subfigure}\hfil 
\begin{subfigure}{0.48\textwidth}
  \includegraphics[width=\linewidth]{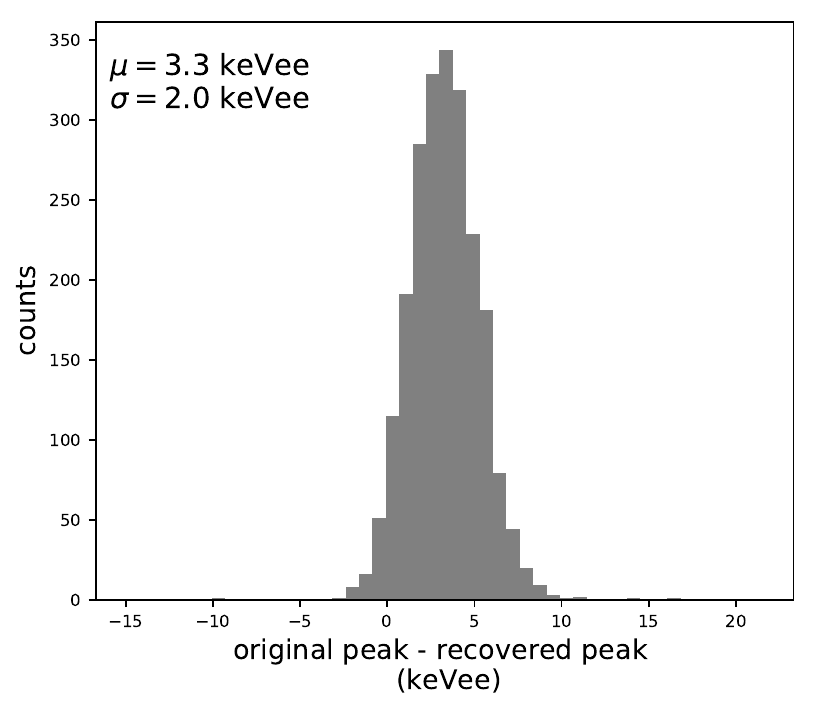}
  \caption{Energy range: 120 - 150 keV}
  \label{fig:250}
\end{subfigure}

\medskip 
\begin{subfigure}{0.48\textwidth}
  \includegraphics[width=\linewidth]{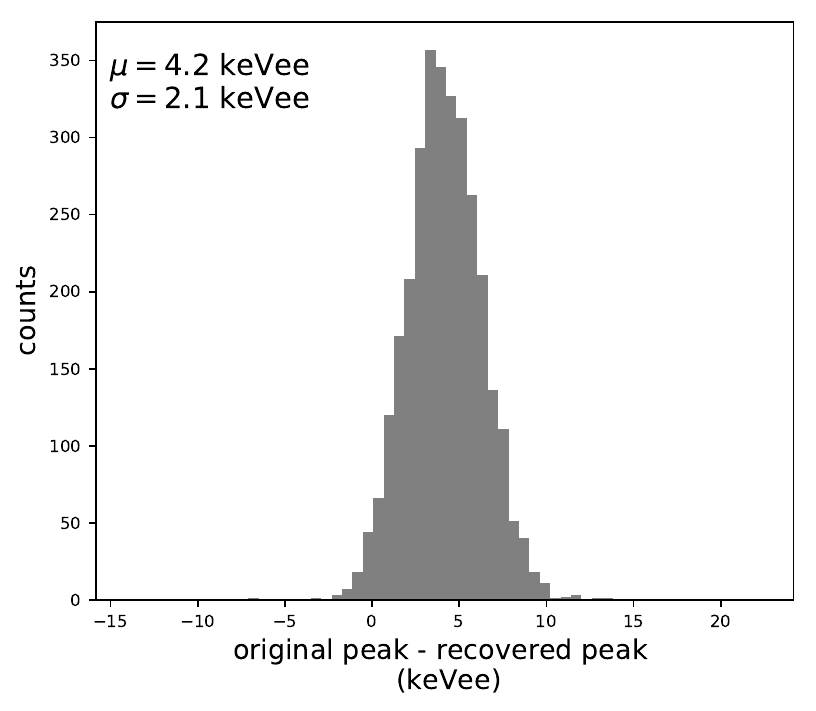}
  \caption{Energy range: 150 - 200 keV}
  \label{fig:350}
\end{subfigure}\hfil
\begin{subfigure}{0.48\textwidth}
  \includegraphics[width=\linewidth]{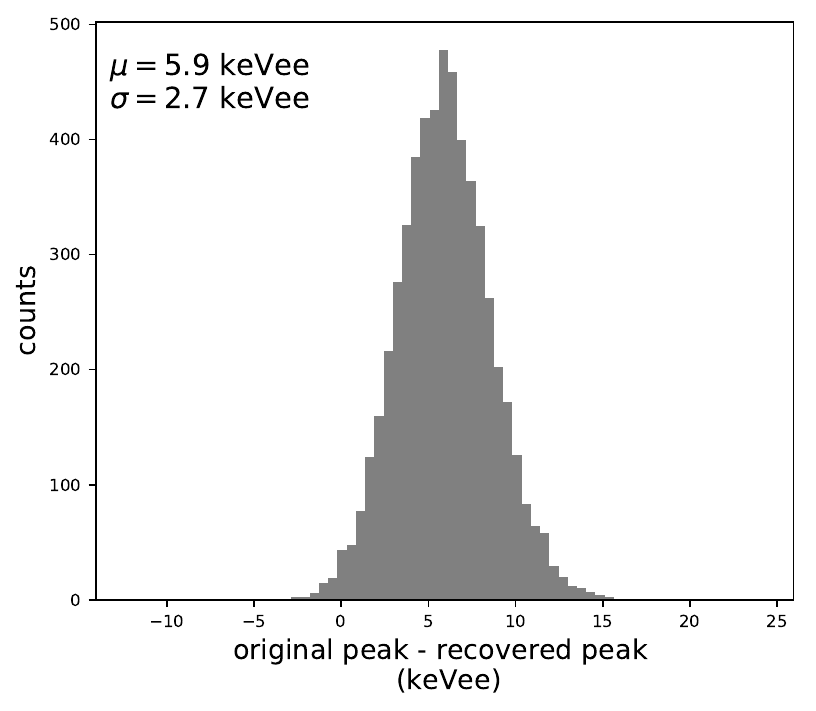}
  \caption{Energy range: 200 - 300 keV}
  \label{fig:450}
\end{subfigure}

\medskip
\begin{subfigure}{0.48\textwidth}
  \includegraphics[width=\linewidth]{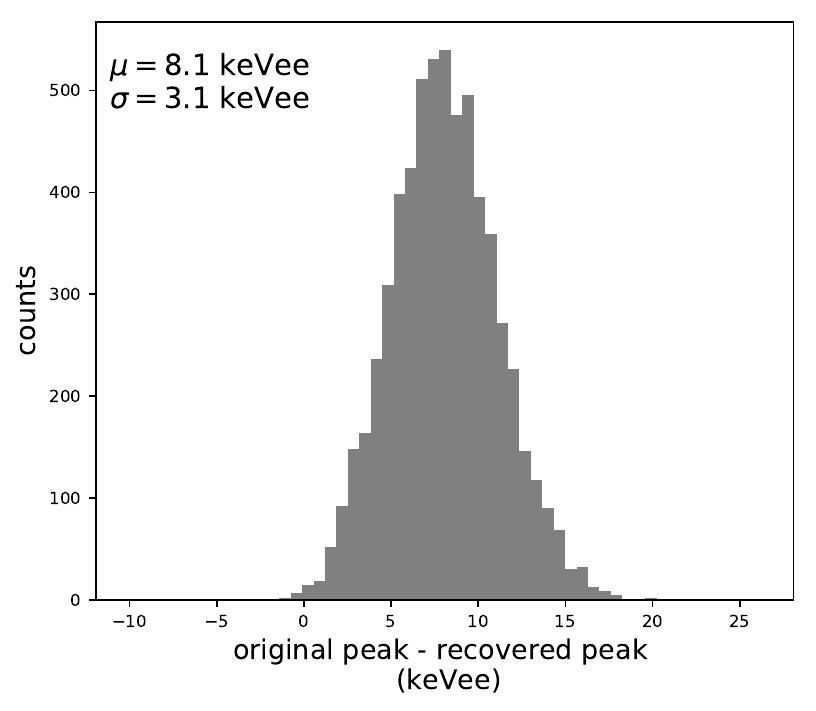}
  \caption{Energy range: 300 - 400 keV}
  \label{fig:550}
\end{subfigure}\hfil 
\begin{subfigure}{0.48\textwidth}
  \includegraphics[width=\linewidth]{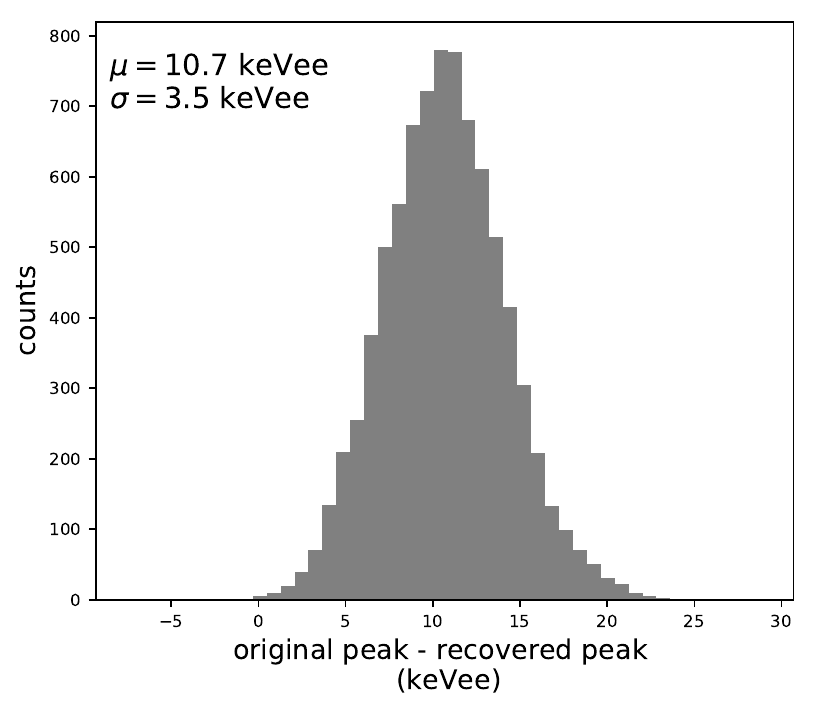}
  \caption{Energy range: 400 - 600 keV}
  \label{fig:650}
\end{subfigure}
\caption{The histograms of the error in energy between the original and the recovered pulse $x_1(n)$ for the energy ranges between 20 to 600 keV using pulse amplitude.}
\label{fig:hist50_eranges}
\end{figure}

\begin{figure}[H]
  \centering
  \begin{subfigure}[!htb]{0.65\linewidth}
    \includegraphics[width=\linewidth]{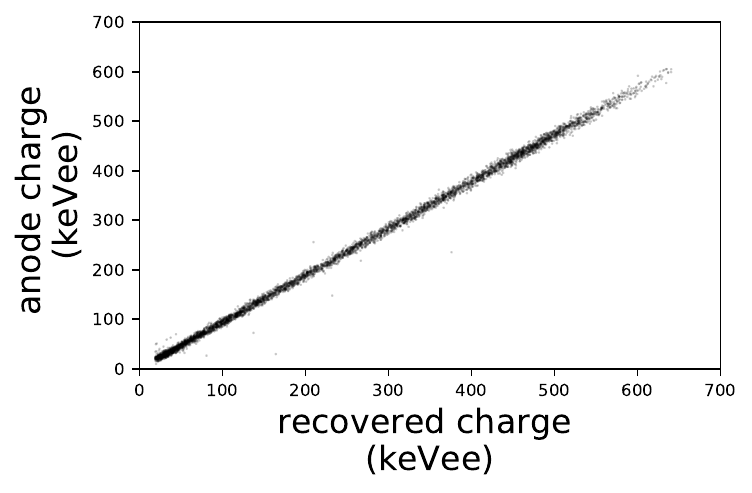}
     
    \caption{Scatter plot between the charge from the second pulse incident on the 9.00 MHz resonator plotted against the recovered charge. }
    \label{fig:ej_ch11p232}
  \end{subfigure}
  \hfill
  \begin{subfigure}[!htb]{0.65\linewidth}
    \includegraphics[width=\linewidth]{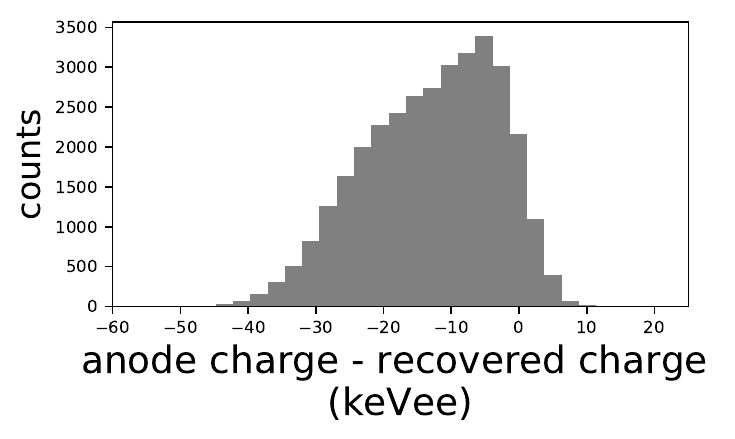}
    
    \caption{The events in Fig. \ref{fig:ej_ch11p232} shown as the distribution of the difference between the charge from the second pulse and the recovered charge with $\sigma$ = 9.8 $\pm$ 0.1 keVee. The error term $\left(\frac{H_{r1}(k)}{H_{r2}(k)}\mathcal{F}\left(\text{tail of } x_1(n)\right)\right)$ introduces a bias and a tail in the distribution with the mean at -10 keVee.}
    \label{fig:ej_ch22p232}
  \end{subfigure}
  \caption{Charge estimation under the second pulse $x_2(n)$ from the recovered pulse $\tilde{x_2}(n)$.}
  \label{fig:ej_chhp232}
\end{figure}

The energy and timing of the original pulses $x_1(n)$ and $x_2(n)$ were again estimated using their recovered counterparts $\tilde{x_1}(n)$ and $\tilde{x_2}(n)$. Because the charge collected under $x_1(n)$ could not be estimated from the partially recovered $\tilde{x_1}(n)$, the pulse amplitude was used to estimate the charge. Fig. \ref{fig:ch50_2d} shows a 2D histogram of the error in energy (or pulse height) as a function of the recovered energy. The mean error in energy increases with recovered energy; this is most likely due to the loss of high frequency components in the recovered signal after applying the low-pass filter, which increase with pulse height. The uncertainty in the error in energy is small for the smaller pulses compared to the larger pulses because the impulse response is not able to recover high frequency components above 160 MHz. The pulses were divided into energy ranges between 90 and 600 keV to show the histograms of the error in energy for different pulse heights (Fig. \ref{fig:hist50_eranges}); the recovered energy within an energy range can be corrected by adding the mean of the corresponding histogram to it. The charge collected under $x_2(n)$ was estimated from the inaccurately recovered pulse $\tilde{x_2}(n)$ with an uncertainty of 9.8 $\pm$ 0.1 keVee shown in Fig. \ref{fig:ej_chhp232}.


\begin{figure}[H]
  \centering
  \begin{subfigure}[!htb]{0.65\linewidth}
    \includegraphics[width=\linewidth]{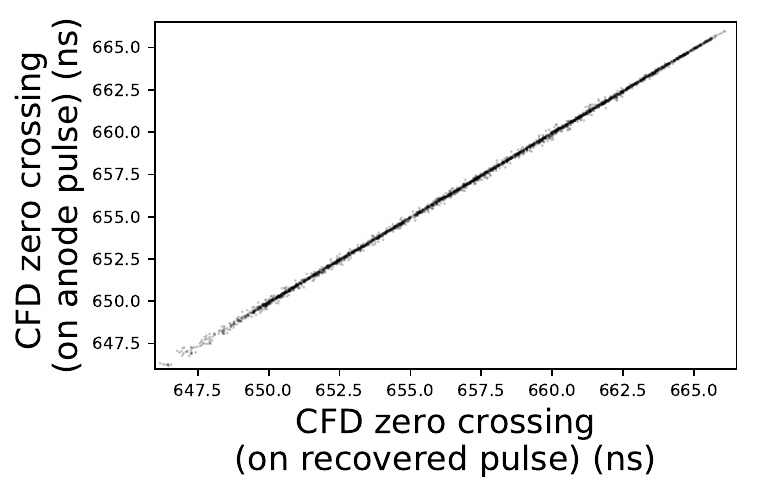}
     
    \caption{Scatter plot between time pick-off of the first pulse incident on the 7.00 MHz resonator plotted against the time pick-off of the recovered pulse.}
    \label{fig:at_rt1132}
  \end{subfigure}
  \hfill
  \begin{subfigure}[!htb]{0.65\linewidth}
    \includegraphics[width=\linewidth]{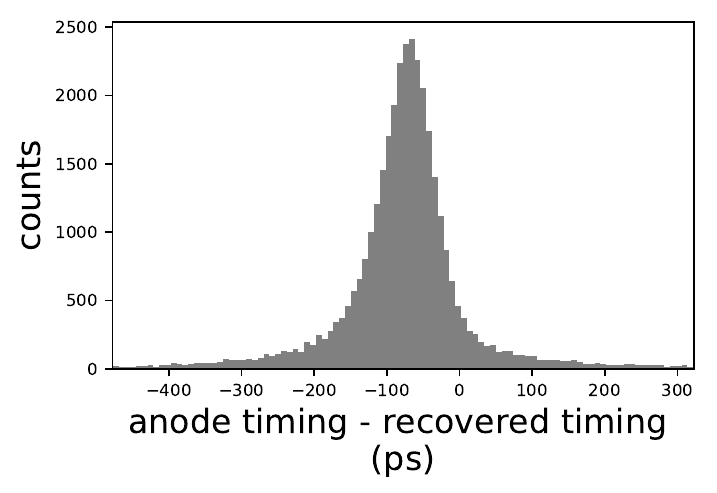}
    
    \caption{The events in Fig. \ref{fig:at_rt1132} shown as the distribution of the difference between the timing of the first pulse and the recovered timing with $\sigma$ = 88 $\pm$ 2.1 ps.The timing was computed by applying CFD with an attenuation fraction of 0.2 and a delay of 6.4 ns. A small bias in the mean is due to the ringing corresponding to the resonant frequency in the additive noise.}
    \label{fig:at_rt2232}
  \end{subfigure}
  \caption{Timing estimation of the first pulse $x_1(n)$ from the recovered pulse $\tilde{x_1}(n)$.}
  \label{fig:at_rtt32}
\end{figure}

\begin{figure}[H]
  \centering
  \begin{subfigure}[!htb]{0.65\linewidth}
    \includegraphics[width=\linewidth]{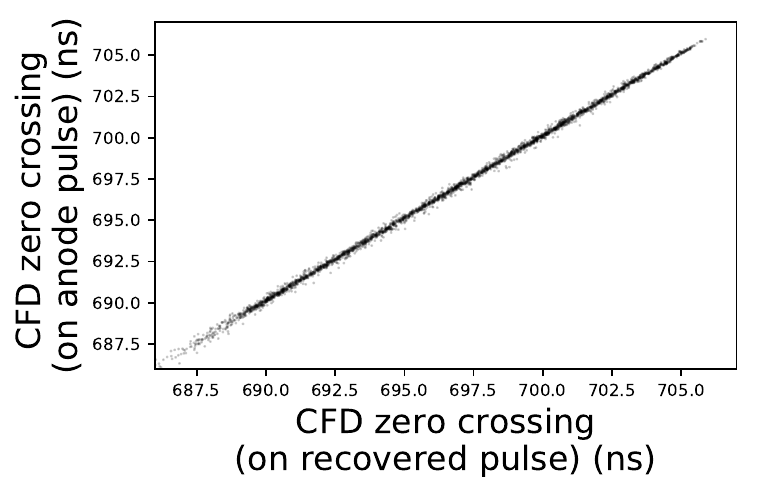}
     
    \caption{Scatter plot between time pick-off of the second pulse incident on the 9.00 MHz resonator plotted against the time pick-off of the recovered pulse.}
    \label{fig:at_rt11p232}
  \end{subfigure}
  \hfill
  \begin{subfigure}[!htb]{0.65\linewidth}
    \includegraphics[width=\linewidth]{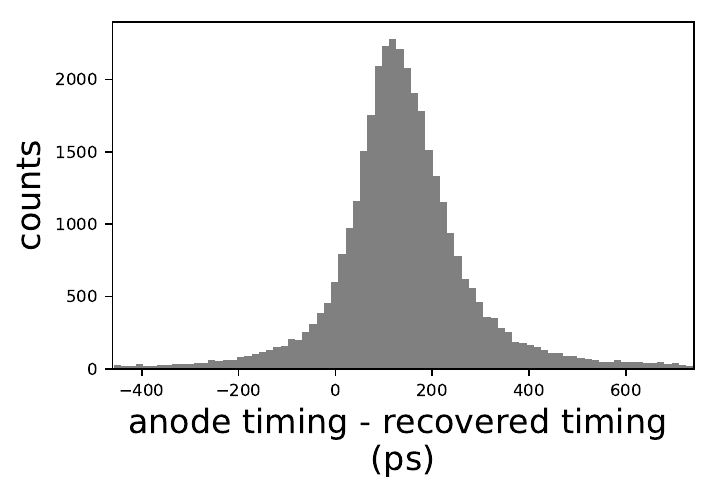}
    
    \caption{The events in Fig. \ref{fig:at_rt11p232} shown as the histogram of the difference between the timing of the second pulse and the recovered timing with $\sigma$ = 141 $\pm$ 2.7 ps. The error term $\left(\frac{H_{r1}(k)}{H_{r2}(k)}\mathcal{F}\left(\text{tail of } x_1(n)\right)\right)$ introduces a positive bias in the distribution with the mean at 177 ps.The timing was computed by applying CFD with an attenuation fraction of 0.3 and a delay of 5.6 ns to account for the attenuation and bandwidth reduction of the pulse due to the passive delay line.}
    \label{fig:at_rt22p232}
  \end{subfigure}
  \caption{Timing estimation of the second pulse $x_2(n)$ from the recovered pulse $\tilde{x_2}(n)$.}
  \label{fig:at_rttp232}
\end{figure}

Fig. \ref{fig:at_rt1132} shows the scatter plot of the time pick-off between the first pulse $x_1(n)$ and the recovered pulse. The uncertainty in the estimate of the timing of the first pulse from the recovered pulse using the histogram in Fig. \ref{fig:at_rt2232} is 88 $\pm$ 2.1 ps, which is the same uncertainty as was observed when the pulses did not overlap. The timing of the second pulse $x_2(n)$ was estimated using the recovered pulse with a much larger uncertainty of 141 $\pm$ 2.7 ps as shown in Fig. \ref{fig:at_rttp232}. 


\section{Noise analysis of the recovered signal}

\begin{figure}[H]
  \begin{subfigure}[t]{0.48\linewidth}
    \includegraphics[width=\linewidth]{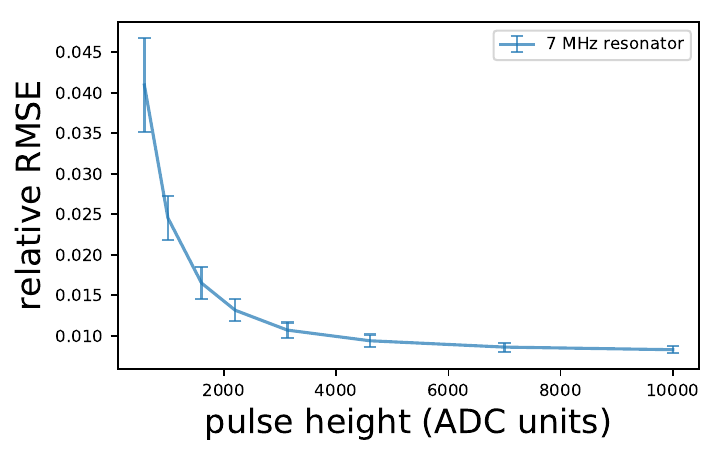}
     
    \caption{The relative RMSE as a function of pulse height for the first pulse incident on the 7 MHz resonator.}
    \label{fig:rmse_7}
  \end{subfigure}
  \hfill
   \hfill
  \begin{subfigure}[t]{0.48\linewidth}
    \includegraphics[width=\linewidth]{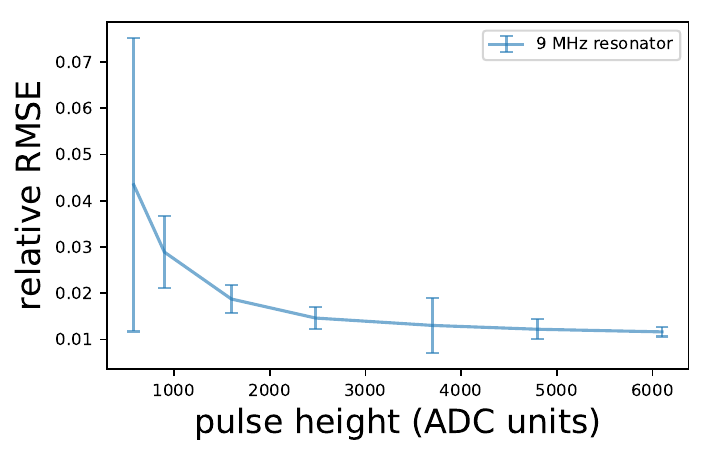}
    
    \caption{The relative RMSE as a function of pulse height for the second pulse incident on the 9 MHz resonator.}
    \label{fig:rmse_9}
  \end{subfigure}
  \caption{The relative RMSE as a function of pulse height.}
  \label{fig:rmse79}
\end{figure}

The residuals (shown in Fig. \ref{fig:reconst1}) for the pulse $x_1(n)$ incident on the 7 MHz resonator remain the same as a function of the pulse height, resulting in a decrease in the relative root-mean-square error (RMSE) with an increase in the pulse height (shown in Fig. \ref{fig:rmse_7}). The residuals (shown in Fig. \ref{fig:reconst2}) for the pulse $x_2(n)$ incident on the 9 MHz resonator increase as a function of the pulse height, resulting in a nearly flat relative RMSE as a function of the pulse height ( shown in Fig. \ref{fig:rmse_9}); the RMSE increases with the pulse height because as the pulse height increases, the decay time of the pulse (to the baseline) also increases resulting in a slight overlap between the two pulses $x_1(n)$ and $x_2(n)$ in time. The overlap increases with pulse height, which also increases the error term (explained in section \ref{overlap}).

\section{Comparison between multiplexing by convolution/deconvolution and \newline modulation/demodulation}
As described in this paper, multiplexing of the detector signals is performed by the convolution between the impulse response of a resonator with input signal in the time domain, which is equivalent to multiplication between them in the frequency domain. Multiplexing by modulation is performed by the multiplication between a carrier current and a detector signal in the time domain, which is equivalent to convolution between them in frequency.     

The deconvolution can theoretically recover the original detector signals from the multiplexed output if the signals do not overlap in time; conversely, demodulation can theoretically recover the detector signals from the multiplexed output if the signals do not overlap in frequency. In case of modulation/demodulation, although the overlap between the tails of the spectra of signals produced by the TESs of adjacent resonant frequencies is reduced by increasing the spacing between the carrier frequencies, it is still present especially for large signals.    

\section{Conclusions}
We tested frequency domain multiplexing of two EJ-309 organic scintillator detectors by convolution and deconvolution when both the detectors produce signals in the same digitizer record. The anode charge and timing of the first pulse was estimated from its recovered counterpart with an uncertainty of about 2.9 keVee and, 87 ps respectively. The anode charge and timing of the second pulse was estimated from the corresponding recovered pulse with a much greater uncertainty of about 7.1 keVee and 137 ps respectively\footnote{The uncertainties depend on the experimental setup used in the measurement: two copies of a EJ-309 pulse from the signal copier acted as inputs to two resonators, where one of the copies was subjected to $\sim$ 35$\%$ attenuation and bandwidth reduction after passing through a passive delay line to delay it by 220 ns.}. Pulse shape discrimination using charge integration performed on both the recovered pulses showed a small decrease in the FOM. 

The reduction in precision in the charge, timing and FOM for the second pulse is attributed to: (1) the attenuation and bandwidth reduction undergone by the second pulse due to the passive delay line; (2) increase in the relative RMSE between the original and the recovered second pulse as a function of the pulse height.

When the two detector pulses overlap in time, the first pulse can be partially recovered from when it arrives until the arrival of the second pulse; the second pulse cannot be recovered accurately; the inaccuracy depends on the error introduced by the portion of the first pulse overlapping the second pulse. No degradation in timing precision was observed for the first pulse.

\section{Future work}
In the future, we plan to perform the multiplexing of silicon photomultiplier (SiPM) fast output signals in the case of double occupancy \cite{SiPM}. As discussed in this paper, even when two detector pulses overlap in time, we can recover both the timing and the peak amplitude of the first pulse; when using a SiPM fast output, the peak amplitude is a good surrogate for charge, so we can estimate the energy of the first pulse as precisely as for the case when the pulses do not overlap in time.

\acknowledgments

This work was sponsored in part by the NNSA Office of Defense Nuclear Nonproliferation R\&D through the Consortium for Verification Technology (CVT), grant number DE-NA0002534.












\end{document}